\documentclass[10pt,reqno]{amsart}
\usepackage{graphicx}
\usepackage{indentfirst,csquotes}
\usepackage{graphicx}
\usepackage{booktabs}
\usepackage{multirow}
\usepackage{amsmath,amssymb,amsfonts}
\usepackage{amsthm}
\usepackage{mathrsfs}
\usepackage{color}
\usepackage{subcaption}
\usepackage{hyperref}
\usepackage{algorithm}
\usepackage{algorithmicx}
\usepackage{algpseudocode}
\usepackage{bm}
\usepackage{array}
\usepackage{enumitem}   

\usepackage{amssymb,amsthm,amsmath}
\usepackage{xcolor,fancyhdr,etoolbox}
\newtheorem{theorem}{Theorem}[]

\newtheorem{lemma}[theorem]{Lemma}
\newtheorem{proposition}[theorem]{Proposition}

\newtheorem*{theorem*}{Theorem} 

\hypersetup{ colorlinks=true, linkcolor=black, filecolor=black, urlcolor=black }

\makeatletter
\renewcommand\section{\@startsection{section}{1}{\z@}%
  {.7\linespacing\@plus\linespacing}{.5\linespacing}%
  {\normalfont\large\bfseries}}
\makeatother

\usepackage{lipsum}

\begin{document}
\title{Quantum Spectral Clustering Framework via Compact Circuit Structures}
\author{Hyeong-gyu Kim\textsuperscript{1}, Siheon Park\textsuperscript{1}, and June-Koo Kevin Rhee\textsuperscript{1,2,*}}
\maketitle

\begin{center}
\small
\textsuperscript{1}School of Electrical Engineering, KAIST, Daejeon, Republic of Korea\\
\textsuperscript{2}Qunova Computing, Inc., Daejeon, Republic of Korea\\[0.6em]
\textit{E-mail:} \href{mailto:khgyu6039@kaist.ac.kr}{khgyu6039@kaist.ac.kr},
\href{mailto:sihoney97@kaist.ac.kr}{sihoney97@kaist.ac.kr},
\href{mailto:rhee.jk@kaist.edu}{rhee.jk@kaist.edu}\\
\textsuperscript{*}Corresponding author
\end{center}
\maketitle

\begin{abstract}
Spectral machine learning methods based on spectral graph theory are powerful but scale poorly due to costly eigen-analysis. Although various eigenvector approximation methods have been proposed, the construction of a partial kernel matrix still remains within their frameworks.
This work presents compact quantum circuit designs for spectral clustering in which the eigenproblem is approximated via a Rayleigh–Ritz formulation that bypasses kernel matrix construction and whose overall depth is dominated by the data-embedding routine.
A rigorous shot complexity analysis is provided, showing that the sampling-based estimation remains tractable for the spectral clustering objective including a penalty term.
Simulations on canonical datasets demonstrate reliable optimization behavior with an under-parameterized hardware-efficient ansatz. Finite-shot simulations further confirm the predicted sampling behavior of the penalty estimator, validating the main claims as a proof of concept.
\end{abstract} 

\bigskip

\section{Introduction}
Spectral graph theory \cite{SpectralGraphChung1997} underpins clustering \cite{SegmentationWeiss1999, NormalizedCutsShi2000, RandomWalksMelia2000, OnSpectralClusteringNg2001}, semi-supervised learning \cite{SemiSupervisedKipf2016}, and manifold methods \cite{LaplacianEigenmapsBelkin2003} by leveraging eigenstructure of graph Laplacians.
In machine learning practice, one builds a kernel graph over $M$ training samples, forms an adjacency/kernel matrix $\bm{A}\in\mathbb{R}^{M\times M}$, and computes quadratic forms or eigenvectors of associated Laplacians.
For dense kernels, the time complexity of constructing and manipulating these objects is typically at least $O(M^2)$, while solving the resulting eigenproblem exactly costs $O(M^3)$.

To mitigate the cost of realization, we make use of the primitives studied in quantum machine learning \cite{VDunjko, JacobB, KFuji} to realize spectral graph theory models by utilizing quantum kernels  \cite{QuantumKernelRealDeviceHavlivcek2019, QuantumKernelSchuld2019} and compact circuit structures (e.g., SWAP-test based similarity estimation) \cite{Mschuld, DKPark, CBlank}.
 Especially, trainable quantum circuits \cite{MBenedetti, ICong, SMangini} can be applied to build an effective learning models with direct evaluation and optimization of the objectives via parameterized quantum circuits.
Motivated by this perspective, we develop a Laplacian-based objective function which is compatible with shallow quantum subroutines while enabling a unified and efficient training-testing workflow.

\begin{figure}[t]
    \centering
    \begin{minipage}[t]{0.9\textwidth}
        \includegraphics[width=\textwidth]{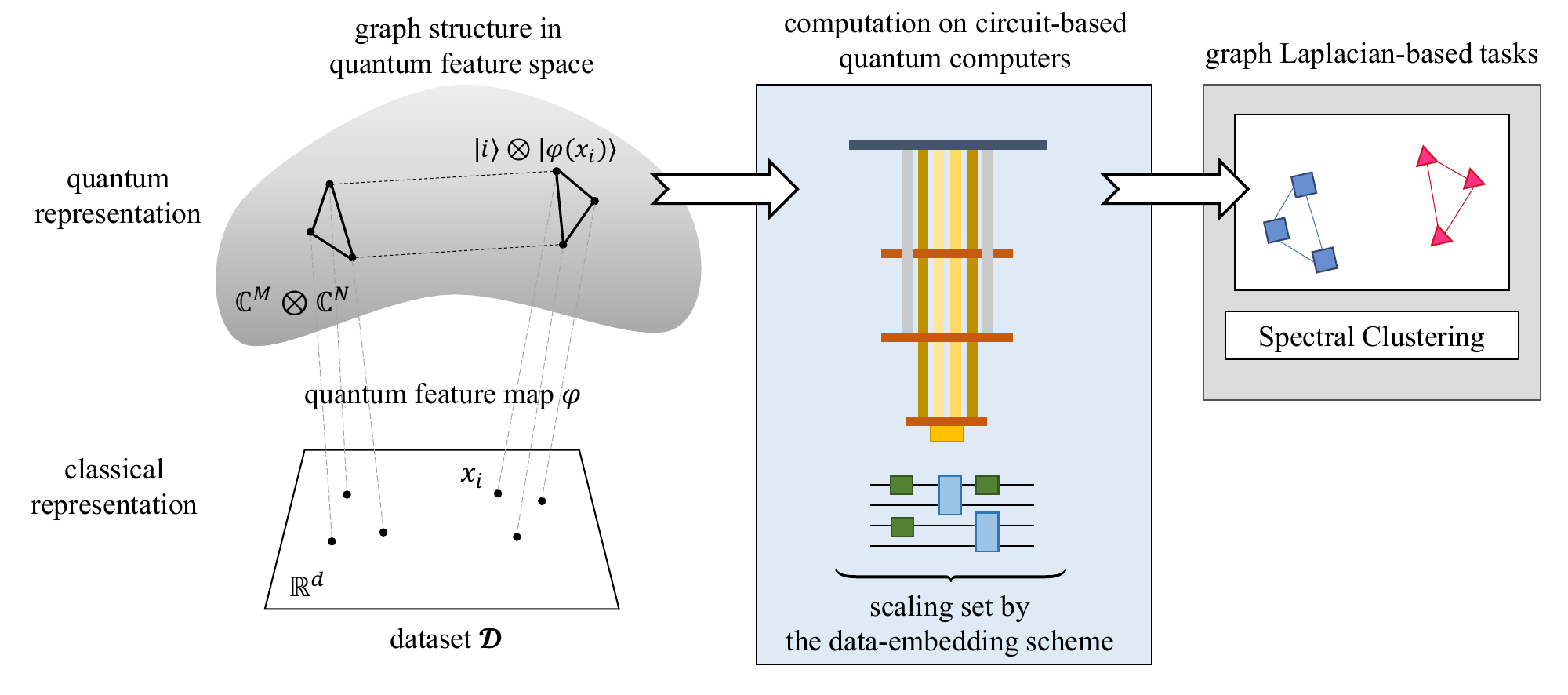}
    \end{minipage}%
    \caption{Overview of processing graph Laplacian-based tasks using quantum kernel methods. Once the data are encoded in Hilbert space, the quadratic form of the graph Laplacian, can be computed with compact quantum circuits on quantum computers, and subsequently applied to learning tasks.}
    \label{fig:overview}
\end{figure}

The broader context of Laplacian-based learning with quantum kernel methods is schematically illustrated in \textbf{Figure \ref{fig:overview}}. Within this framework, we focus on variational approximate spectral clustering driven by Laplacian quadratic forms, estimated via compact quantum circuit structures.
Our contributions are:

\begin{itemize}
\item \textbf{Circuit structures for Laplacian quadratic forms.}
We present a compact family of circuits that estimate three quantities induced by the fidelity kernel: (i) an adjacency quadratic form, (ii) a degree-weighted quadratic form, and (iii) a projector penalty. Building on these, we realize - through these circuits - a normalized Rayleigh-quotient objective adapted from classical spectral clustering and weighted kernel PCA \cite{Suyken_SVMPCA, Suykens_WPCA}, whose projector term suppresses the trivial eigenvector, together with a score function that assigns cluster membership.

\item \textbf{Complexity and shot-noise analysis.}
We rigorously analyze the total training-time complexity using concentration inequalities, accounting for statistical noise introduced by a finite number of measurement shots. This analysis indicates that the quantity estimated by the penalty circuit can be estimated reliably with a relatively small number of shots.
\end{itemize}

The remainder of this paper is organized as follows. Section~\ref{sec:methods} introduces the proposed variational spectral learning objective and compact circuit estimators for Laplacian-induced quadratic forms, along with the score function used for inference on unseen data. Section~\ref{sec:analysis} provides circuit-depth scaling and a finite-shot sample-complexity analysis, with particular emphasis on the penalty estimator. Section~\ref{sec:results} reports simulation results on canonical datasets, and Section~\ref{sec:conclusions} concludes with implications and directions for future work.

\newpage
\section{Methods}\label{sec:methods}
\subsection{Preliminaries and notation}
\begin{itemize}
\item \textbf{Graph Laplacian and spectral clustering}

Given a dataset $\mathcal{D}=\{\bm{x_0},\dots,\bm{x_{M-1}}\}\subset\mathbb{R}^d$, let $\kappa(\cdot,\cdot)$ be a nonnegative similarity.
Define
\begin{align}
(\bm{A})_{ij} &= \kappa(\bm{x_i},\bm{x_j}), \\
(\bm{D})_{ij} &= \delta_{ij}\sum_{k=0}^{M-1}\kappa(\bm{x_i},\bm{x_k}), \\
 \bm{L}_{\mathrm{sym}} &= \bm{I} - \bm{D}^{-1/2}\bm{AD}^{-1/2}. \label{eq:symmetric_laplacian}
\end{align}
For given symmetric Laplacian matrix $\bm{L_{\text{sym}}}$, the clustering task can be done by solving a relaxed trace minimization whose solutions are eigenvectors of $\bm{L_{\text{sym}}}$.
For binary clustering (two-way cut), the second smallest eigenvector of $\bm{L}_{\mathrm{sym}}$ often suffices.
A multiway clustering can also be obtained by recursively applying the binary spectral cut to split clusters in a hierarchical/divisive manner.

\item \textbf{Quantum feature map and fidelity kernel}

Let a feature map $\varphi$ be implemented via a data-dependent unitary $U_{\varphi(\bm{x})}$:
\begin{align}
U_{\varphi(\bm{x})}\lvert 0\rangle^{\otimes n} = \lvert \varphi(\bm{x})\rangle
\end{align}
where $n$ stands for the number of qubits used for quantum feature map. We use the fidelity kernel
\begin{align}
\kappa(\bm{x_i},\bm{x_j}) := \big|\langle \varphi(\bm{x_i})\mid \varphi(\bm{x_j})\rangle\big|^2.
\end{align}
This kernel naturally arises from SWAP-test style estimators \cite{SPark}.
\item \textbf{Data embedding / indexed state preparation}

We use an indexed data embedding unitary $U_{\varphi,\mathcal{D}}$ that prepares
\begin{align}
U_{\varphi,\mathcal{D}}\Big(\lvert +\rangle^{\otimes m} \otimes \lvert 0\rangle^{\otimes n}\Big)
= \frac{1}{\sqrt{M}}\sum_{i=0}^{M-1} \lvert i\rangle \otimes \lvert \varphi(\bm{x_i})\rangle,
\end{align}
with $m=\lceil\log_2 M\rceil$ and $n$ denoting the number of qubits used by the quantum feature map $U_{\varphi(\bm{x})}$.
We assume the use of standard uniformly-controlled single-qubit gates for the implementation \cite{UniformlyControlledGate}, which requires $O(M)$ single-qubit gates and controlled-NOT gates, together with a single diagonal gate acting on $O(\log(M))$ qubits \cite{UniformlyControlledGate2}.

\item \textbf{Trainable weighting state}

We define a trainable weighting state
\begin{align}
\lvert \alpha_\theta\rangle := U_\alpha(\bm{\theta})\lvert 0\rangle^{\otimes m}
= \sum_{i=0}^{M-1}\alpha_i(\bm{\theta})\lvert i\rangle,
\qquad \|\bm{\alpha(\theta)}\|_2=1.
\end{align}
We choose a hardware-efficient ansatz with $O(\mathrm{poly}\log M)$ parameters, but our method is agnostic to the specific ansatz.

\end{itemize}

\subsection{Objective function and estimators}

\begin{figure}[t]
    \centering
    \begin{minipage}[t]{0.45\textwidth}
        \centering
        \includegraphics[width=\textwidth]{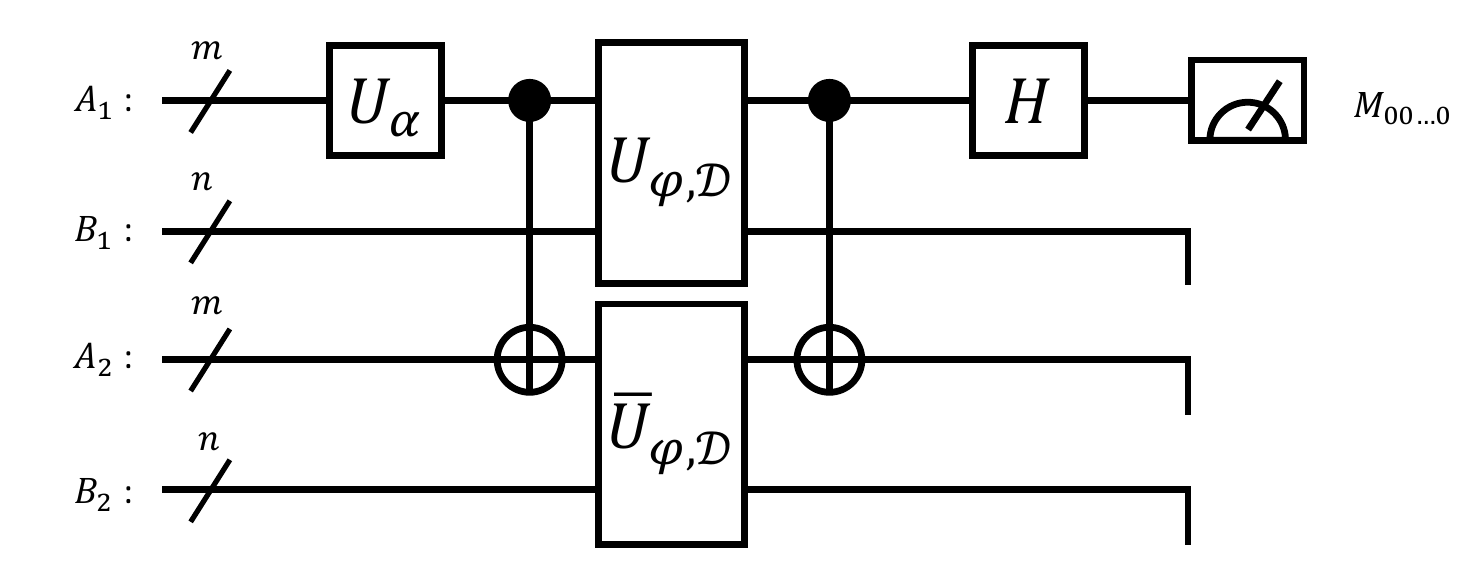}
        \subcaption{$q_a(\alpha)$}
        \label{fig:circuits-a}
    \end{minipage}%
    \hspace{0.1cm} 
    \begin{minipage}[t]{0.45\textwidth}
        \centering
        \includegraphics[width=\textwidth]{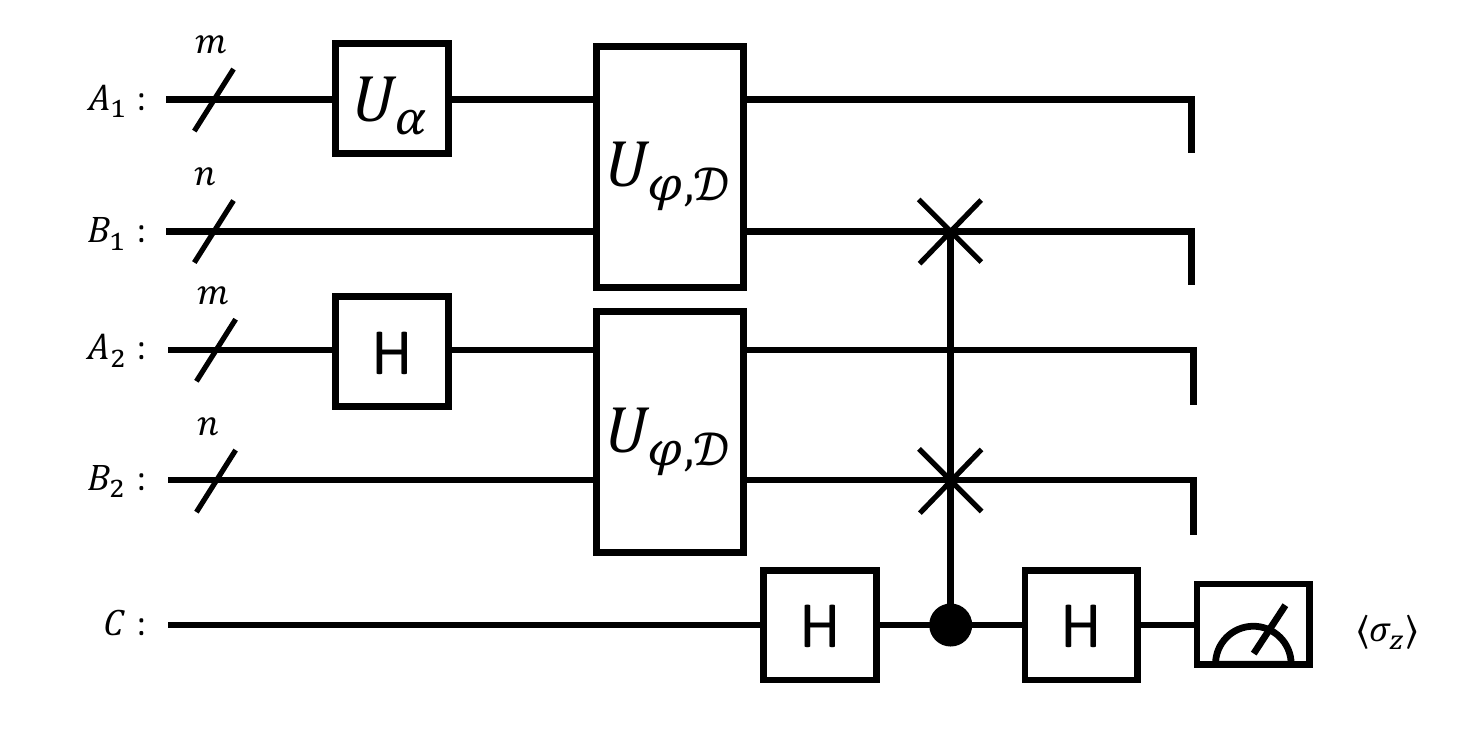}
        \subcaption{$q_d(\alpha)$}
        \label{fig:circuits-b}
    \end{minipage}%
    \hspace{0.1cm} 
    \begin{minipage}[t]{0.45\textwidth}
        \centering
        \includegraphics[width=\textwidth]{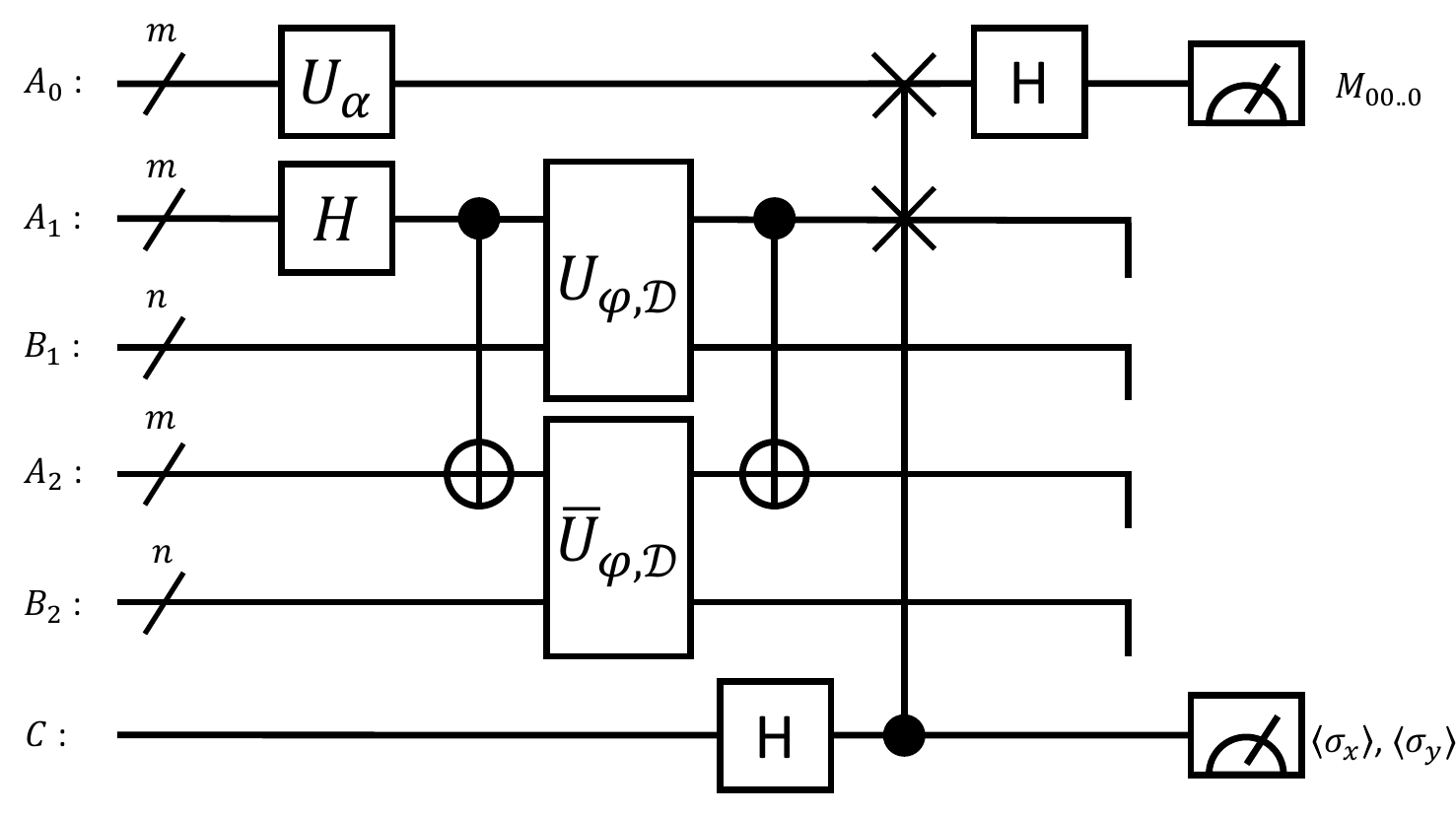}
        \subcaption{$q_{p}(\alpha)$}
        \label{fig:circuits-c}
    \end{minipage}
    \begin{minipage}[t]{0.45\textwidth}
        \centering
        \includegraphics[width=\textwidth]{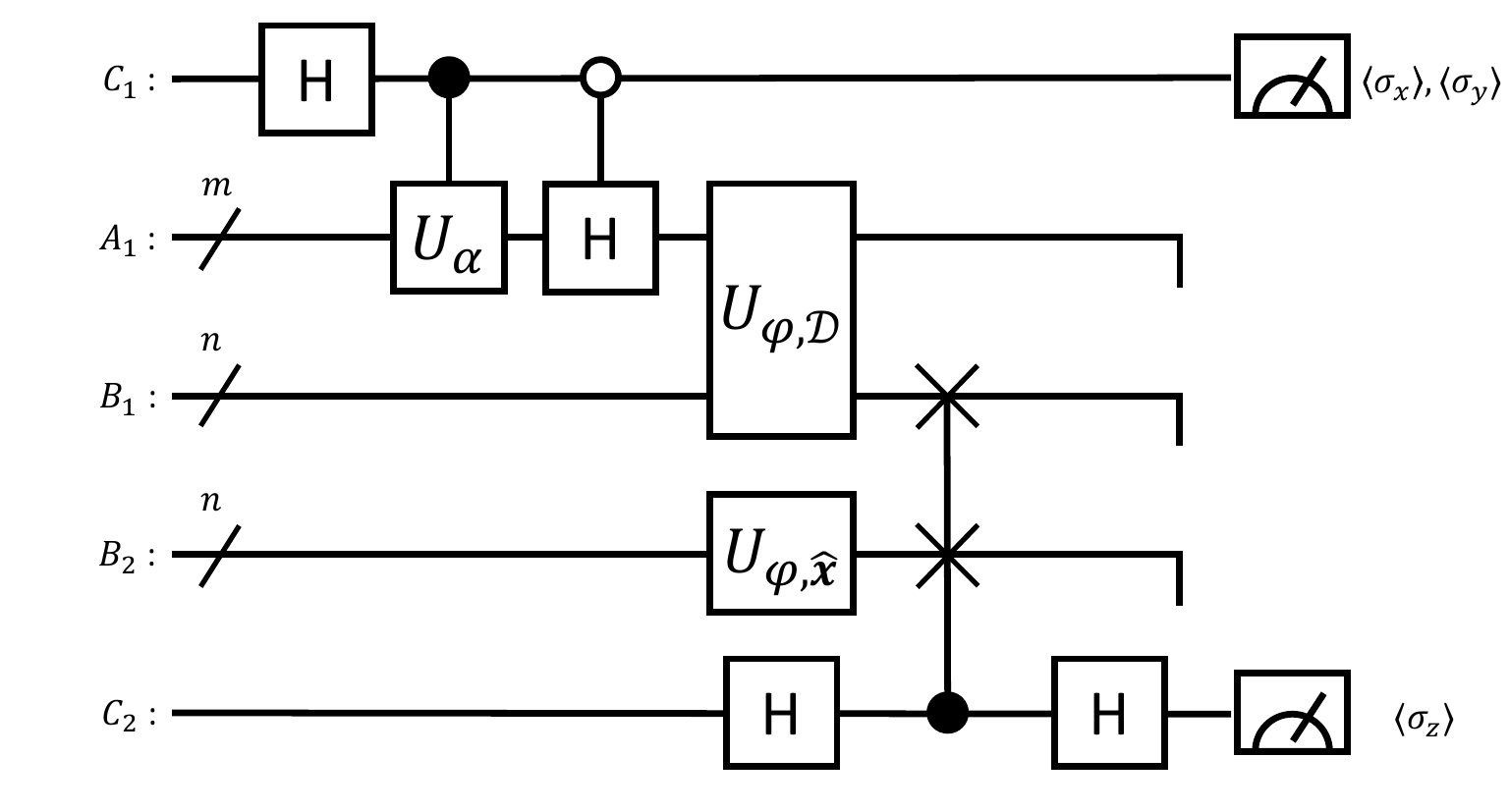}
        \subcaption{$q_{H_x}, q_{H_y}$}
        \label{fig:circuits-d}
    \end{minipage}
    \caption{Quantum circuits corresponding to the estimators proposed in main text. $\overline{U}_{\varphi, \mathcal{D}}$ denotes the complex conjugate of $U_{\varphi, \mathcal{D}}$. The operator $U_{\varphi, \hat{x}}$ is a test data embedding unitary that encodes $\hat{x}$ via $U_{\varphi, \hat{x}} |0\rangle^{\otimes n} = |\bm{\hat{x}}\rangle$.}
    \label{fig:circuits}
\end{figure}

We target a normalized-Laplacian objective using $\bm{\alpha}$, which can be derived directly from the quadratic form of the symmetric Laplacian in Equation \ref{eq:symmetric_laplacian} in the Supplementary material:
\begin{align}\label{eq:objective-main}
\max_{\bm{\alpha=\alpha(\theta)}} \quad
\mathcal{J}(\bm{\alpha}) :=
\frac{\bm{\alpha^\dagger A\alpha} - \xi \bm{\alpha^\dagger D \vec{1}\vec{1}^\dagger D\alpha}}{\bm{\alpha^\dagger D \alpha}},
\qquad \xi>0,
\end{align}
where $\bm{\vec{1}}$ denotes the all-ones vector. \textbf{Figure \ref{fig:circuits}} summarizes the quantum circuits used to estimate the terms appearing in the objective Equation \ref{eq:objective-main}.
The key point is that $\mathcal{J}(\bm{\alpha})$ can be evaluated by estimating three scalar terms. We define estimators $q_a(\bm{\alpha})$, $q_d(\bm{\alpha})$, and $q_p(\bm{\alpha})$ such that
\begin{align}
q_a(\bm{\alpha}) &:= \frac{1}{M}\sum_{i,j}\alpha_i\bar{\alpha}_j\,\kappa(\bm{x_i},\bm{x_j})=\frac{1}{M}\bm{\alpha^\dagger A\alpha}, \label{eq:q_a}\\
q_d(\bm{\alpha}) &:= \frac{1}{M}\sum_{i}|\alpha_i|^2\sum_j \kappa(\bm{x_i},\bm{x_j})=\frac{1}{M}\bm{\alpha^\dagger D\alpha}, \label{eq:q_d}\\
q_p(\bm{\alpha}) &:= \frac{1}{M^2}\Big|\sum_i \bar{\alpha}_i\, d_i\Big|^2 = \frac{1}{M^2} \bm{\alpha^\dagger D\, \vec{1}\vec{1}^\dagger D\,\alpha},
\qquad d_i:=\sum_j \kappa(\bm{x_i},\bm{x_j}). \label{eq:q_p}
\end{align}
Each estimator can be obtained with trivial classical post-processing from the circuits in \textbf{Figure \ref{fig:circuits-a}}, \textbf{\ref{fig:circuits-b}}, and \textbf{\ref{fig:circuits-c}}, respectively. Implementation details are provided in Appendix \ref{app:estimators}.

For inference on unseen data, we additionally introduce a score function estimator. 
Given a trained $\bm{\hat{\alpha}}$ and a test point $\bm{\hat{x}}$, we define
\begin{align}
s(\bm{\hat{x}}, \bm{\hat{\alpha}}) &:= \sum_{i=0}^{M-1}\hat{\alpha}_i\,\kappa(\bm{x_i},\bm{\hat{x}}), \\
\hat{\phi}(\bm{\hat{x}}, \bm{\hat{\alpha}}) &:= \mathrm{phase}\,s(\bm{\hat{x}})
= \arctan2\!\left(q_{H_y}, q_{H_x}\right) \in [-\pi,\pi],
\end{align}
where
\begin{align}
q_{H_x} &:= \sum_{i=0}^{M-1} |\hat{\alpha}_i| \cos(\phi_i)\,|\langle \bm{x_i} \mid \bm{\hat{x}}\rangle|^2, \\
q_{H_y} &:= \sum_{i=0}^{M-1} |\hat{\alpha}_i| \sin(\phi_i)\,|\langle \bm{x_i} \mid \bm{\hat{x}}\rangle|^2,
\end{align}
and $\phi_i$ denotes the phase of $\hat{\alpha}_i$. Operationally, the quantities $q_{H_x}$ and $q_{H_y}$ are obtained from the same circuit in \textbf{Figure \ref{fig:circuits-d}} by measuring the ancilla qubit $C_1$ in the $\sigma_x$ and $\sigma_y$ bases, respectively (i.e., $q_{H_x}=\langle \sigma_x\rangle_{C_1}$ and $q_{H_y}=\langle \sigma_y\rangle_{C_1}$).
The phase $\hat{\phi}(\bm{\hat{x}}, \bm{\hat{\alpha}})$ can be converted into cluster labels via thresholding or circular clustering.
We provide a detailed explanation of the corresponding circuit \textbf{Figure \ref{fig:circuits-d}} in Appendix~\ref{app:test-circuits}.

\begin{figure}[t]
\centering
\includegraphics[width=0.85\textwidth]{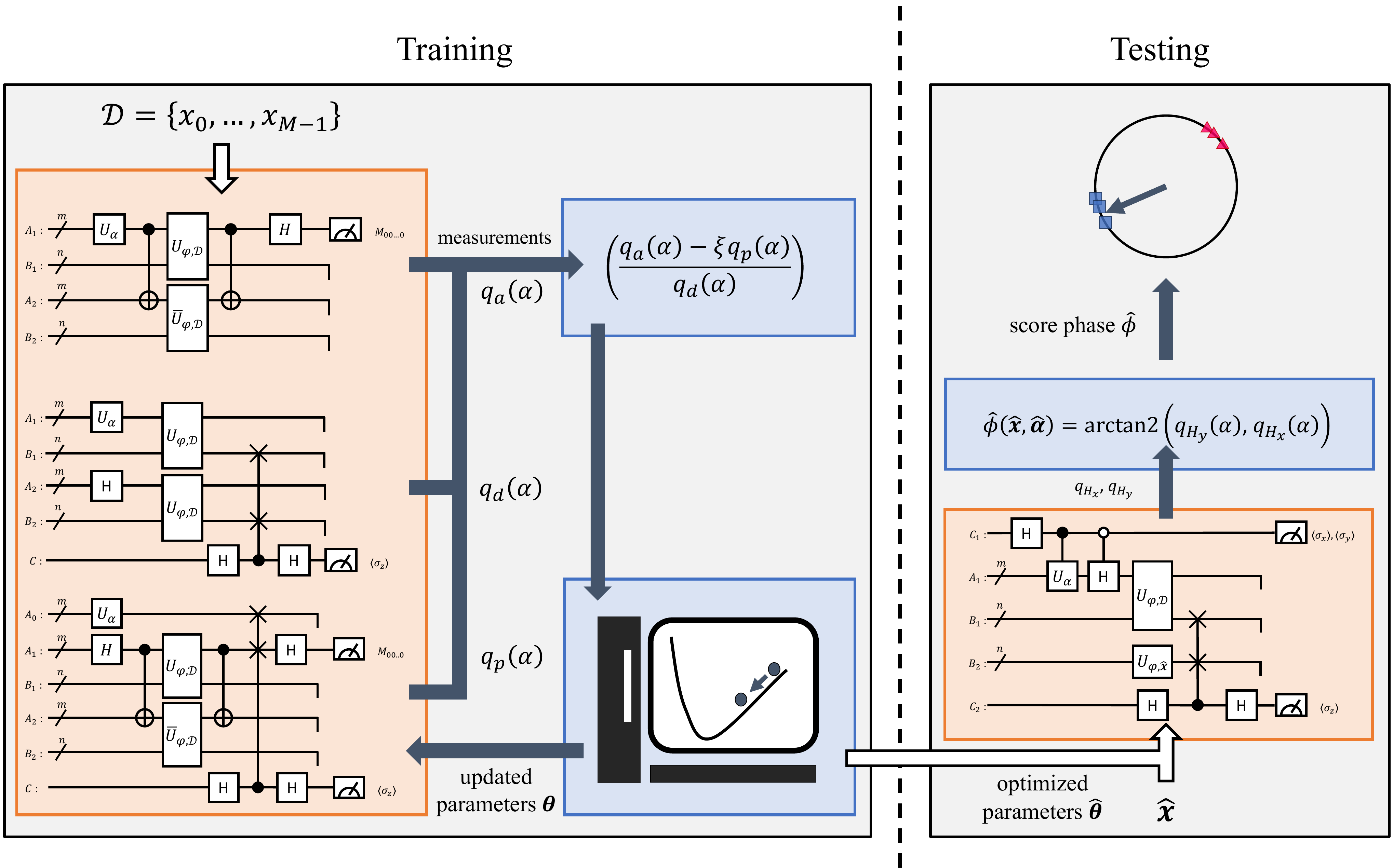}
\caption{High-level workflow. Orange: quantum subroutines for estimating $(q_a,q_d,q_p)$ and test-time scoring. Blue: classical optimizer and post-processing.}
\label{fig:workflow}
\end{figure}

To place these components in context, \textbf{Figure~\ref{fig:workflow}} summarizes the end-to-end pipeline: the objective is optimized using the estimator circuits during training, and the learned $\hat{\alpha}$ is subsequently used for scoring and label assignment via the phase $\hat{\phi}$.

\section{Theoretical Analysis}\label{sec:analysis}
\subsection{Circuit Depth and Shot Complexity}
Since the circuits in \textbf{Figure \ref{fig:circuits}} share the same overall structure, it suffices to focus on three components: the data-embedding unitary $U_{\varphi,\mathcal{D}}$, the weighting unitary $U_\alpha$, and the SWAP-test subroutine.
In general, the depth of $U_{\varphi,\mathcal D}$ dominates the overall circuit depth. This block is implemented via uniformly controlled gates, and its scaling depends on the choice of embedding scheme.
For example, if we restrict the embedding scheme to elementary gate sets whose depth scales linearly with $M$, then quantum feature maps with all-to-all connectivity in the embedding circuit incur a depth of $O(Mn^2)$, whereas those restricted to nearest-neighbor connectivity incur $O(Mn)$ \cite{SPark}.
Accordingly, each estimator circuit has overall depth
\begin{align}
C_{\mathcal{D}}(M,n) + C_\alpha(M) + O(n + \log M).
\end{align}
Here, the leading $C_{\mathcal{D}}(M,n)$ term accounts for the indexed data-embedding block $U_{\varphi,\mathcal D}$, the term $C_\alpha(M)$ is the depth required to prepare the trainable weighting state $|\alpha_\theta\rangle$ on the $m=\lceil\log_2 M\rceil$-qubit index register, and the additive $O(n+\log M)$ term collects the controlled-SWAP subroutines: those in $q_d(\bm\alpha)$, $q_{H_x}$, and $q_{H_y}$ act on the $n$-qubit data registers, whereas the one in $q_p(\bm\alpha)$ acts on the $m$-qubit index register.

A recurring concern is that $q_p(\bm{\alpha})$ (Equation \ref{eq:q_p}) has smaller  magnitude scale compared to $q_a(\bm{\alpha})$ (Equation \ref{eq:q_a}) and $q_d(\bm{\alpha})$ (Equation \ref{eq:q_d}), which could suggest a larger shot requirement.
To make this scale issue explicit, we introduce the auxiliary quantity
\begin{align}\label{eqn:Z_orig}
Z(\bm{\alpha}) := \frac{1}{M^2}\,|\langle \bm{d},\bm{\alpha}\rangle|\,|\langle \bm{\vec{1}},\bm{\alpha}\rangle|,
\qquad \|\bm{d}\|_2\le M^{3/2},
\end{align}
which corresponds to the value estimated by the quantum circuit corresponding to \textbf{Figure \ref{fig:circuits-c}}.
(Equation \ref{eqn:Z_orig} corresponds to the result of the estimator circuit in \textbf{Figure \ref{fig:circuits-c}} before the post-processing. See Appendix \ref{app:estimators} for more details).

In what follows, we show that the number of samples required to estimate $Z(\bm{\alpha})$ to a prescribed tolerance scales favorably: for a target failure probability $\delta$, the required shot count grows at most quasi-polynomially in $1/\delta$.

\begin{theorem}[Informal, shot complexity of the penalty]
Let $\{Z_i\}_{i=0}^{r-1}$ be i.i.d.\ copies of
$Z(\bm{\alpha})$, $\bm\alpha\sim\mathrm{Unif}(S^{2M-1})$, with mean $\mu:=\mathbb E[Z]$. For sufficiently small $\delta\in(0,1)$, some $c>1$, and suppose the precision parameter satisfies
$$
s \in \Omega \left(\sqrt{\frac{\ln(1/\delta)}{\delta}}
\exp \left (-\frac{1}{2} \ln^{c}(1/\delta)\right)\right).
$$
Then, to guarantee
$$
\Pr\!\left(\Big|\frac1r\sum_{i=0}^{r-1}Z_i-\mu\Big|\le\frac{s}{M}\right)\ge 1-\delta,
$$
it suffices to take
$$
r = O \left(\delta \exp(\ln^{c}(1/\delta))\right).
$$
\end{theorem}

A more detailed derivation is provided in Appendix \ref{app:shot-proof-sketch}.

For the Rayleigh-quotient objective, gradient descent requires $O\!\left(\frac{1}{\Delta}\log\frac{M}{\varepsilon_{\mathrm{GD}}}\right)$ iterations to reach an $\varepsilon_{\mathrm{GD}}$-accurate neighborhood \cite{RayleighConvergence}, where $\Delta := 1 - \lambda_{1}/\lambda_{0}$ denotes the spectral gap for the leading eigenvector, with $\lambda_0 > \lambda_1$ the two largest eigenvalues.
Using the parameter-shift rule \cite{ParamShift1,ParamShift2}, each gradient evaluation incurs an additional multiplicative cost proportional to the number of trainable parameters, which we assume scales as $O(\mathrm{poly}\log M)$.

We identify the statistical estimation error with the optimization tolerance, i.e., $\varepsilon \equiv \varepsilon_{\mathrm{GD}}$.
Combining (i) the per-evaluation circuit depth dominated by state preparation $C_{\mathcal{D}}(M,n)$, (ii) the parameter-shift overhead $O(\mathrm{poly}\log M)$ and convergence ratio of the Rayleigh-quotient objective, (iii) the standard $1/\varepsilon^2$ sampling factor for estimating expectation values to accuracy $\varepsilon$, and (iv) the (at most) multiplicative overhead required for the penalty term, the total runtime to obtain a trained circuit for $\alpha$ scales as
\begin{align}
O\left(\frac{C_{\mathcal{D}}(M,n)}{\Delta\,\varepsilon^{2}} \mathrm{poly}\log(M,1/\varepsilon)\cdot\max\left\{1,\delta \exp(\ln^{c}(1/\delta))\right\}\right).
\end{align}
(Here we drop the subscript on $\delta$ for brevity.)

\subsection{Comparison with Classical Approximation Methods}\label{sec:comparison}

Table~\ref{tab:comparison} compares the proposed method with the naive dense baseline and three families of scalable classical eigen-approximation. Our framework adopts the quantum fidelity kernel as the underlying similarity measure, following standard practice in the quantum kernel methods literature \cite{QuantumKernelRealDeviceHavlivcek2019, QuantumKernelSchuld2019}.

\begin{enumerate}[label=(\roman*)]
    \item \textbf{Low-rank approximation}.
The Nystr\"om method selects $s$ landmark samples from $M$ training points and constructs a low-rank factorization of the kernel matrix, reducing the cost of the subsequent eigen-analysis to $O(Ms^2 + s^3)$~\cite{Nystrom}. Here, the $O(s^3)$ term arises from eigendecomposing the $s \times s$ landmark block, while the $O(Ms^2)$ term comes from expanding the resulting eigenvectors back to all $M$ training points. The required $s$ depends on the spectral decay of the chosen kernel: kernels with rapidly decaying eigenvalues admit small $s$. Although Nystr\"om avoids constructing the full kernel matrix, it still requires explicit access to the $M \times s$ submatrix between landmark and remaining samples.
    \item \textbf{Random feature approximation}.
Random Fourier Features (RFF) \cite{RFF} construct $D$ explicit features per sample in $O(MD)$ time, from which a $D \times D$ feature covariance matrix is assembled at $O(MD^2)$ cost and subsequently diagonalized at $O(D^3)$ cost, yielding a total of $O(MD^2 + D^3)$. Although RFF avoids constructing the full $M \times M$ kernel matrix, it is applicable only to shift-invariant kernels $\kappa (\bm{x}, \bm{y}) = \kappa (\bm{x} - \bm{y})$. The quantum fidelity kernel $|\langle \varphi(\bm{x}) | \varphi(\bm{y})\rangle|^2$ is not shift-invariant in general, and RFF therefore does not constitute a direct alternative to our method. RFF and the proposed approach address distinct kernel classes rather than competing on the same one.
    \item \textbf{Iterative eigensolvers}.
The Lanczos method \cite{Lanczos, golub2013matrix} computes the top-$k$ eigenpairs through repeated matrix-vector products $\bm{v} \mapsto \bm{M}\bm{v}$, needing no explicit matrix entries, and is especially efficient when the matrix is sparse. The fidelity kernel, however, is not assumed sparse and its entries must be estimated rather than read at unit cost, so each matrix-vector product requires $O(M^2)$ estimates, reinstating dense kernel access despite Lanczos being matrix-free.
\end{enumerate}

Methods (i) and (iii) still require explicit access to a (partial or full) kernel matrix. When that kernel is the quantum fidelity kernel, individual entries are no longer available at unit cost: each must be estimated through a quantum subroutine such as the SWAP test, and controlling the cumulative Frobenius-norm error of the estimated matrix to accuracy $\epsilon$ requires $O(\epsilon^{-2} M^4)$ shots for the full $M \times M$ matrix and $O(\epsilon^{-2} M^2 s^2)$ shots for the $M \times s$ submatrix used by Nystr\"om-type schemes \cite{QuantumKernelRealDeviceHavlivcek2019}. This is a property of entry-wise access, and it does not arise for estimators that interact with the kernel only through aggregate quadratic forms, such as those in Section \ref{sec:methods}.

In classical settings, the methods in (i)-(iii) can clearly outperform the na\"ive dense pipeline and in many regimes perform eigen-analysis more efficiently than the algorithm proposed here. The picture shifts only once one commits to the quantum fidelity kernel, where the entry-wise query workload can dominate the rest of the computation. Our method therefore complements rather than supersedes classical eigen-approximation, targeting the regime where the underlying kernel is itself quantum and kernel-query cost becomes a primary concern.

An empirical comparison against the Nystr\"om and Lanczos baselines, on the same datasets and splits used in Section~\ref{sec:results}, is reported in the Supplementary Material.

\begin{table*}[t]
\centering
\caption{Characterization of representative eigen-approximation strategies for spectral learning. Methods are contrasted by eigensolver family, kernel-access pattern, and asymptotic scaling within each method's own cost model. Here $\tilde{O}(\cdot)$ suppresses polylogarithmic factors in $M$ and $1/\epsilon$, $C_\mathcal{D}:=C_\mathcal{D}(M,n)$ is the depth of the indexed data-embedding block $U_{\varphi,\mathcal{D}}$, $M$ is the number of training samples, $s$ the Nystr\"om landmark count, $D$ the number of random Fourier features, $t$ the Lanczos iteration count, and $\Delta$ the relevant spectral gap.}
\label{tab:comparison}
\renewcommand{\arraystretch}{1.4}
\setlength{\tabcolsep}{6pt}
\small
\begin{tabular}{@{}
  >{\raggedright\arraybackslash}p{1.6cm}
  >{\raggedright\arraybackslash}p{2.6cm}
  >{\raggedright\arraybackslash}p{3.5cm}
  >{\raggedright\arraybackslash}p{2.9cm}
  >{\raggedright\arraybackslash}p{3.5cm}@{}}
\toprule
\textbf{Method}
& \textbf{Eigensolver family}
& \textbf{Kernel access pattern}
& \textbf{Scaling}
& \textbf{Key dependence / caveat} \\
\midrule
Na\"ive dense
& Direct dense diagonalization
& Full $M\times M$ kernel matrix
& $O(M^3)$
& Exact reference; returns the full eigenbasis. \\
\midrule
Nystr\"om \cite{Nystrom}
& Low-rank landmark approximation
& $M\times s$ submatrix (landmarks vs.\ rest)
& $O(Ms^2+s^3)$
& $s$ set by the spectral decay of the kernel. \\
\midrule
RFF \cite{RFF}
& Random-feature approximation
& Explicit $M\times D$ features; no kernel matrix
& $O(MD^2+D^3)$
& Standard RFF assumes shift-invariance; not directly applicable to the fidelity kernel. \\
\midrule
Lanczos \cite{Lanczos, golub2013matrix}
& Krylov-subspace iteration
& Matrix-free ($\bm{v}\mapsto\bm{M}\bm{v}$); dense kernel forces $O(M^2)$ entries per product
& $O(t\,M^2)$ \mbox{(dense kernel)}
& $t$ set by the targeted spectral gap; favorable when the kernel is sparse. \\
\midrule
\textbf{Ours}
& Variational Rayleigh-quotient optimization
& Aggregate quadratic forms only; entries never queried individually
& $\tilde{O}\!\big(C_\mathcal{D}(M,n)/(\Delta\epsilon^{2})\big)$
& Dependent on the data-embedding circuit depth $C_\mathcal{D}$.\\
\bottomrule
\end{tabular}
\end{table*}

\section{Results}\label{sec:results}
\begin{figure}[t]
    \centering
    \begin{minipage}[t]{0.25\textwidth}
        \includegraphics[width=\textwidth]{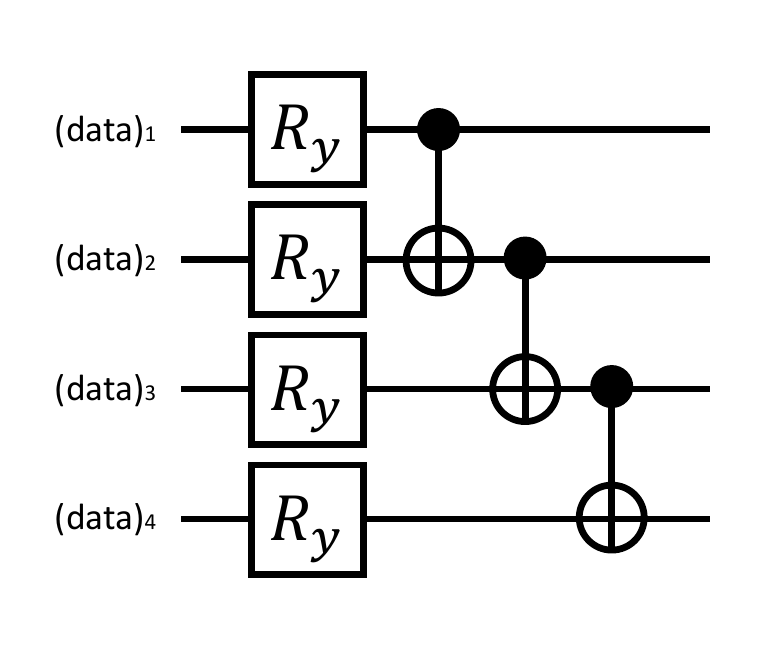}
        \subcaption{}
        \label{fig:simulations-embedding (U)}
    \end{minipage}%
    \hspace{0.1cm} 
    \begin{minipage}[t]{0.30\textwidth}
        \includegraphics[width=\textwidth]{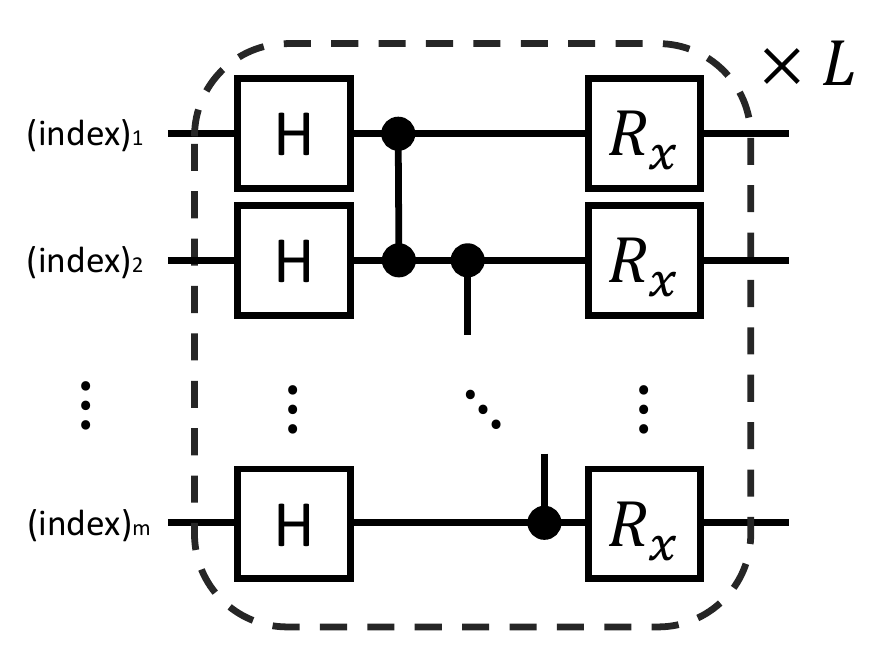}
        \subcaption{}
        \label{fig:simulations-PQC (W)}
    \end{minipage}%
    \hspace{0.1cm}
    \caption{Circuits used in the simulations. 
(a) Representative data-encoding block for $U_{\varphi,\mathcal{D}}$ (single-qubit $R_y$ rotations followed by a nearest-neighbor entangling chain). 
(b) Hardware-efficient ansatz for the weighting unitary $U_{\alpha}(\theta)$ on the $m$-qubit index register, consisting of layerwise Hadamards, entangling gates, and parameterized $R_x$ rotations repeated for $L$ layers.}
    \label{fig:simulation circuits}
\end{figure}

\begin{figure}[t!]
    \centering
    \begin{minipage}[t]{0.40\textwidth}
        \includegraphics[width=\textwidth]{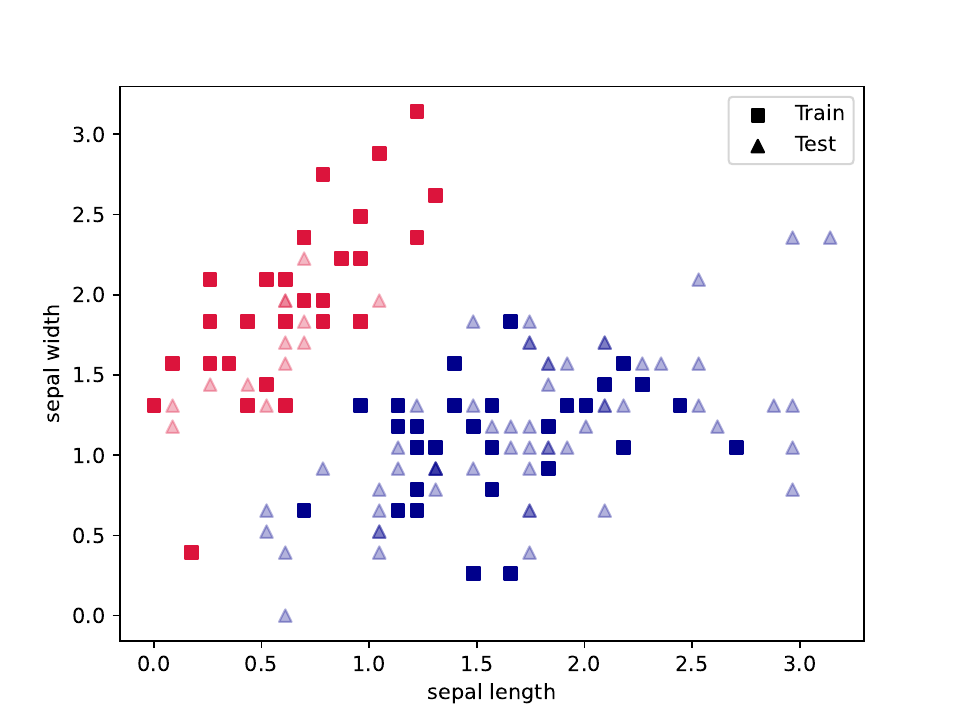}
        \subcaption{}
        \label{fig:iris_visualization}
    \end{minipage}
    \begin{minipage}[t]{0.30\textwidth}
        \includegraphics[width=\textwidth]{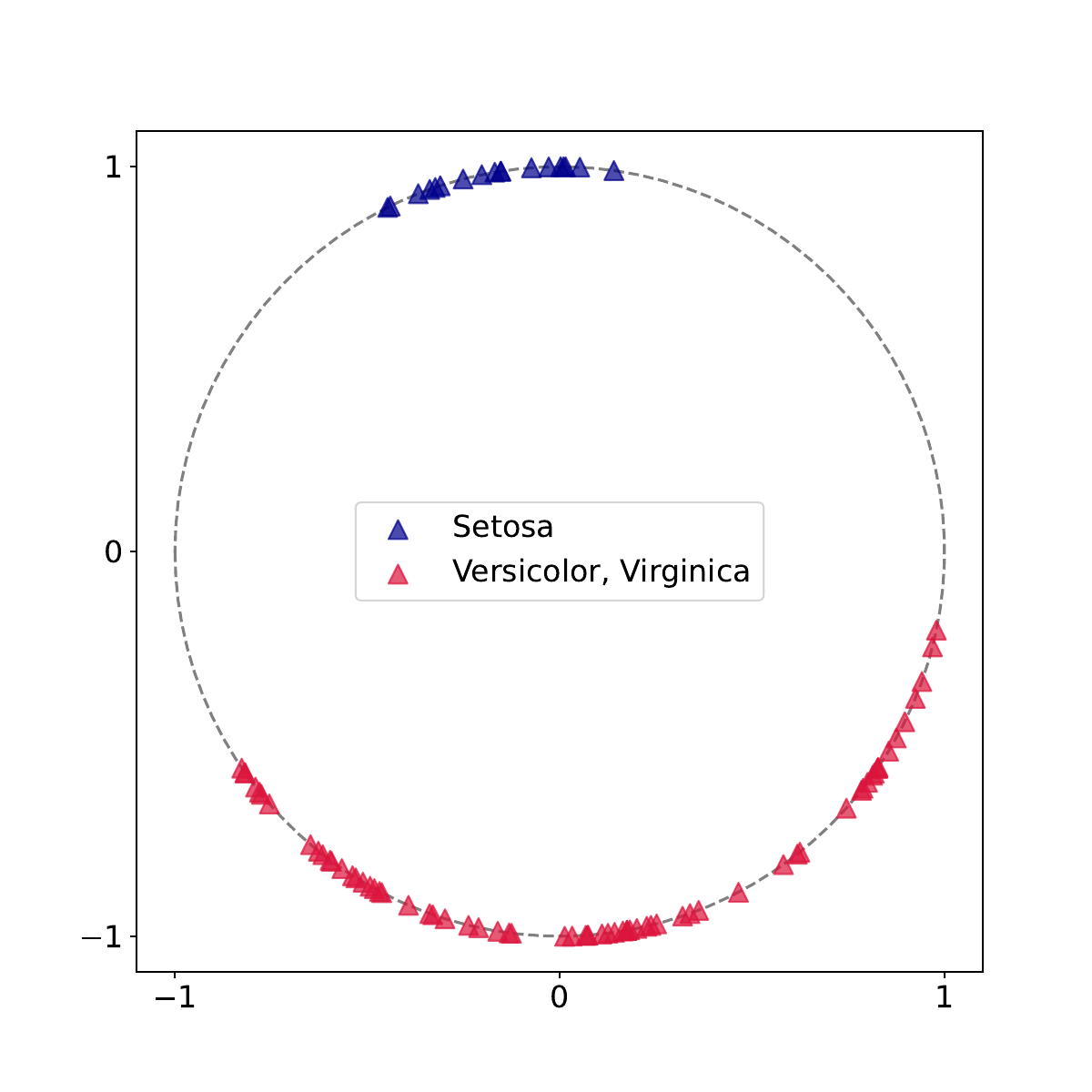}
        \subcaption{}
        \label{fig:iris_test_example}
    \end{minipage}
    \caption{Binary clustering example on the \textit{Iris} dataset. 
(a) 2D visualization (sepal length vs. sepal width) of the samples used in this run; squares denote training points and triangles denote held-out test points. Colors indicate the ground-truth classes and are shown for visualization only. 
(b) Test scores produced by the trained model, visualized on the unit circle using the phase 
$\hat{\phi}(\hat{x},\hat{\alpha})$; each test point is plotted at $(\cos\hat{\phi},\sin\hat{\phi})$. 
Cluster labels are obtained by thresholding (or circular clustering) on $\hat{\phi}$.}
    \label{fig:iris_example}
\end{figure}

Throughout this section, as described earlier, we encode the training data by applying a uniformly controlled gate between the index register and the data register to construct the encoding operator $U_{\varphi, \mathcal{D}}$, where the encoding block is shown in \textbf{Figure \ref{fig:simulations-embedding (U)}}. For the weighting unitary $U_{\alpha}$, we use a hardware-efficient ansatz proposed in prior work \cite{SukinS}, illustrated in \textbf{Figure \ref{fig:simulations-PQC (W)}}. We perform the simulations using the PennyLane quantum simulation framework \cite{Pennylane}.

\begin{figure}[t]
    \centering
    \begin{minipage}[t]{0.49\textwidth}
        \vspace{0pt} 
        \centering
        \includegraphics[width=0.9\textwidth]{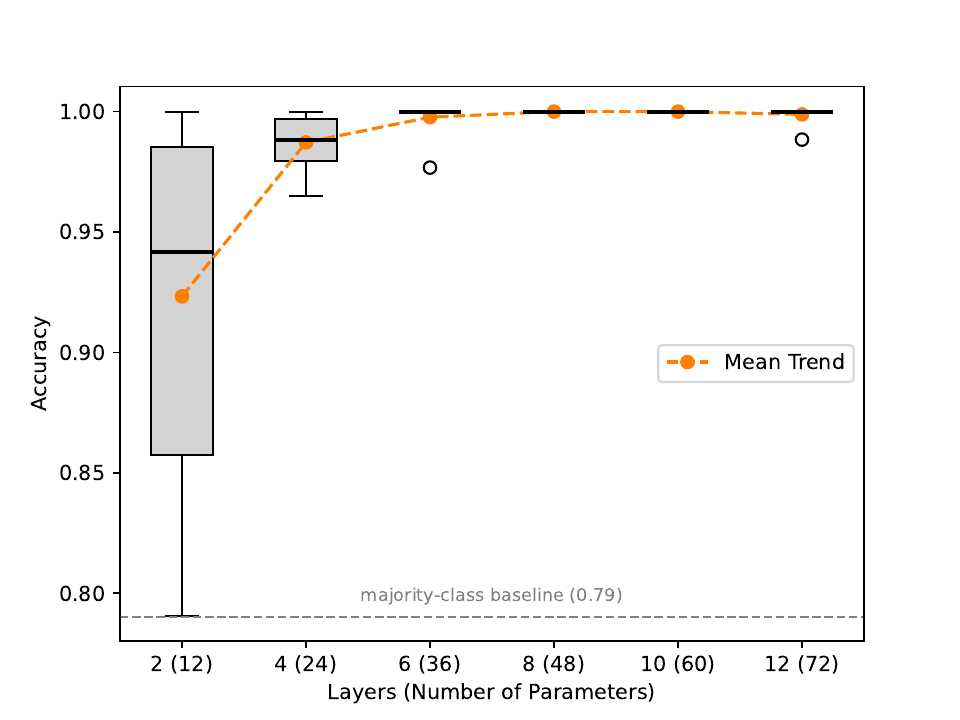}
        \subcaption{}
        \label{fig:iris_10folds}
    \end{minipage}\hfill
    \begin{minipage}[t]{0.49\textwidth}
        \vspace{0pt} 
        \centering
        \includegraphics[width=0.9\textwidth]{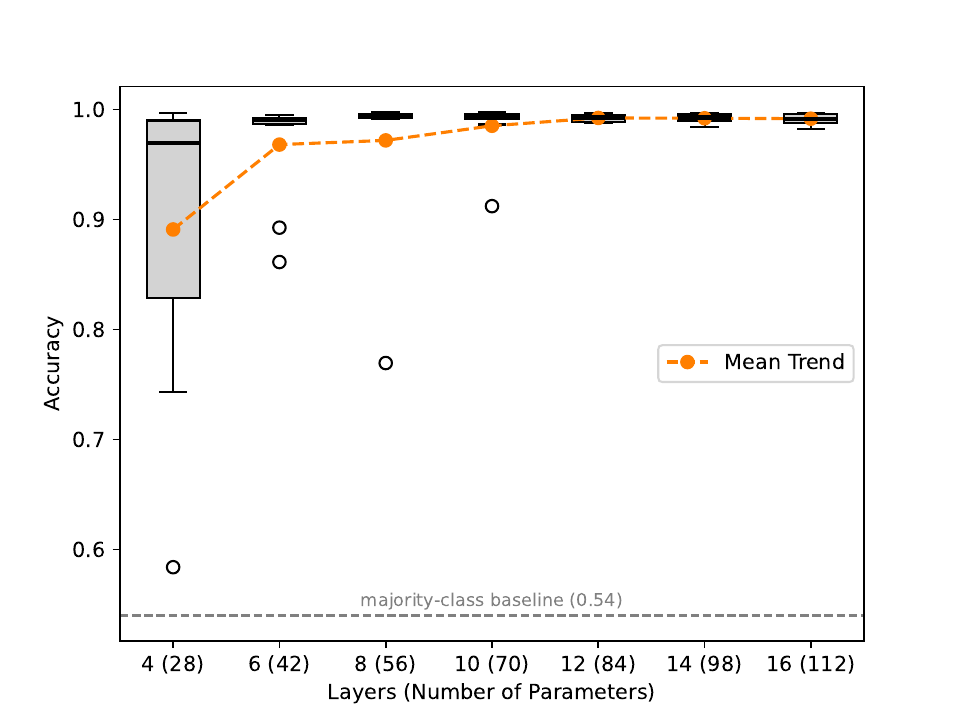}
        \subcaption{}
        \label{fig:MNIST_10folds}
    \end{minipage}\hfill
    \caption{Test accuracy versus circuit depth (number of layers $L$) for (a) the \textit{Iris} and (b) the \textit{MNIST} dataset. 
For each $L$, the box plot and open circles summarize accuracy statistics (median, interquartile range, whiskers) and outliers, respectively, over 10 independent train/test splits. The dashed curves indicate the mean trend.}
    \label{fig:10folds}
\end{figure}

\begin{figure}[t!]
    \centering
    \begin{minipage}[t]{0.35\textwidth}
        \vspace{0pt} 
        \centering
        \includegraphics[width=\linewidth]{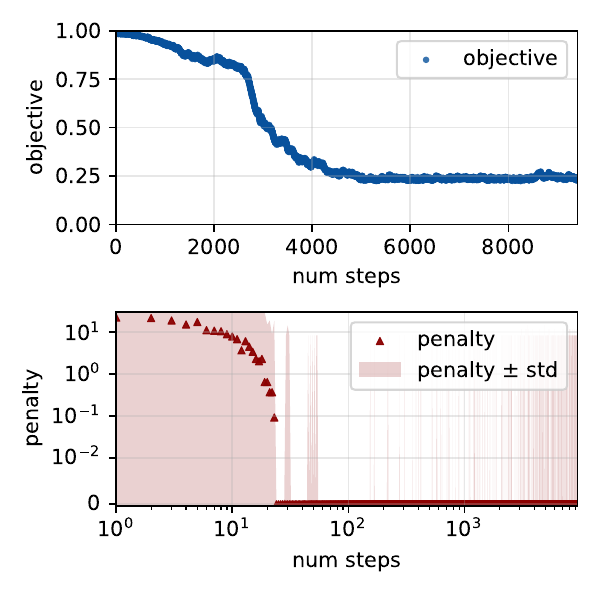}
        \subcaption{}
        \label{fig:penalty_ribbon}
    \end{minipage}\hfill
    \begin{minipage}[t]{0.60\textwidth}
        \vspace{0pt} 
        \centering
        \includegraphics[width=0.49\linewidth]{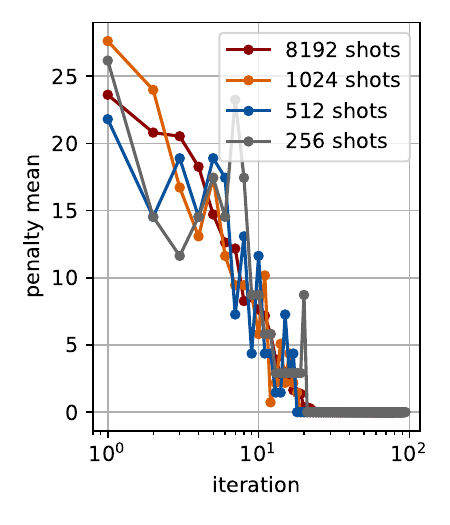}
        \includegraphics[width=0.49\linewidth]{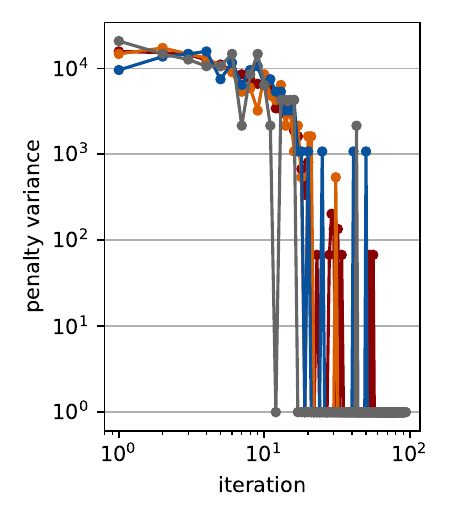}
        \subcaption{}
        \label{fig:penalty_finite_shots}
    \end{minipage}\hfill
    \caption{Finite-shot behavior of the penalty estimator on the \textit{Iris} dataset ($L=6$). (a) Training trajectories of the objective and the penalty term under $R=8192$ shots. (b) Comparison of the sample mean (left) and sample variance (right) of the penalty estimator for different number of shots over the first 100 iterations, illustrating that reliable penalty estimation does not require an excessively large number of shots.}
    \label{fig:shot_simulations}
\end{figure}

As training datasets for spectral clustering, we use the \textit{Iris} and \textit{MNIST} datasets from Scikit-learn \cite{Scikit} python package. The \textit{Iris} dataset has four attributes describing iris flower and is divided into three classes. For this simulation, only two classes—\{\textit{Setosa}\} and \{\textit{Versicolor, Virginica}\}—are considered for clustering. The \textit{MNIST} dataset consists of handwritten digit images of 28×28 pixels. In our simulation, only the classes corresponding to the digits ‘\textit{0}’ and ‘\textit{1}’ are used to evaluate performance. The original 784-dimensional image inputs are reduced to four principal components via principal component analysis (PCA), so that both datasets share a common four-dimensional input representation, and the quantum encoding accordingly operates on a four-qubit data register throughout our simulations. For both datasets, each of the four features is then min-max rescaled to the range $[0, \pi]$ using statistics computed on the training split only (and applied identically to the test split to avoid data leakage). The rescaled features are encoded as rotation angles of single-qubit $R_y$ gates - one feature per qubit - following the embedding circuit shown in 
\textbf{Figure \ref{fig:simulations-embedding (U)}}.

\textbf{Figure \ref{fig:iris_example}} illustrates a representative single run on the \textit{Iris} dataset, visualizing the sampled training/test points and the resulting phase scores on the unit circle. \textbf{Figure \ref{fig:10folds}} summarizes the test accuracy as a function of circuit depth for the \textit{Iris} and \textit{MNIST} datasets, obtained using a noiseless quantum simulator.
For the \textit{Iris} dataset (\textbf{Figure \ref{fig:iris_10folds}}), 64 samples were selected for training and the remaining 86 samples were used for testing. 
This train/test split procedure was repeated 10 times with different random splits (training on the selected subset and testing on the remainder) to assess robustness.
The results indicate that, despite employing a relatively small parameterized quantum circuit with 24 parameters - corresponding to $L=4$ layers of the ansatz in \textbf{Figure~\ref{fig:simulations-PQC (W)}} - the model achieves an average test accuracy of 98.7\%. 
Increasing the number of layers improves the average accuracy and yields more stable training with fewer outliers; notably, for $L \ge 6$, the average accuracy stabilizes at or above 99.8\%. 
This behavior is consistent with our design objective of keeping the circuit depths short in $M$ while maintaining clustering quality on these datasets.

For the \textit{MNIST} dataset (\textbf{Figure \ref{fig:MNIST_10folds}}), 128 samples were selected for training and 1,024 samples were randomly chosen from the remaining data for testing (with the same 10-split evaluation protocol). 
The results indicate that reliable performance begins around $L=6$ layers, and from $L=8$ onward the accuracy stabilizes at an average of 97.2\%. 
Furthermore, as $L$ increases, the standard deviation decreases and outliers become less frequent, indicating that the proposed objective can be optimized stably at the small circuit depths used here.

We also examined how the variance converges to zero as the objective function is optimized on a finite-shot simulator.
Using the same \textit{Iris} dataset, we introduced a robustness heuristic that uses the penalty term to guide the optimizer toward the optimum while having negligible impact on overall optimization performance. 
Details are provided in Appendix~\ref{app:robustness-heuristic}.

\textbf{Figure~\ref{fig:penalty_ribbon}} summarizes the finite-shot behavior of the penalty estimator during training. 
As optimization proceeds, the sample mean of the penalty $q_p(\bm{\alpha})$ drops rapidly, and the corresponding sample variance contracts and approaches zero as the mean becomes negligible. 
This trend is consistent with the concentration-based assumptions used in our Bernstein-inequality analysis and empirically supports the lemmas in Appendix~\ref{app:shot-proof-sketch}.

We also verified that the number of shots required to compute the penalty does not generally need to be large. In \textbf{Figure \ref{fig:penalty_finite_shots}}, we simulated the computation of the penalty term for different numbers of shots over the first $10^2$ iterations. The simulation results indicate that the variance of the penalty tends to converge at zero less stably when the number of shots is small, whereas the mean of the penalty tends to converge to zero within about $10$ iterations for all four choices of the shot number. This indicates that, for the purpose of computing the penalty in our formulation, the required number of shots does not need to be critically large relative to that used for the other terms in the objective function, which is consistent with the theorem proposed in the main text.

In summary, these experiments serve as a proof of concept on canonical
datasets: they demonstrate that the proposed objective can be optimized
reliably with parameterized quantum circuits, and that the estimator
circuits - including the penalty estimator - behave as predicted by our
analysis. Moreover, the finite-shot experiments indicate that the penalty estimator is well-behaved: its mean rapidly decreases during training and its variance contracts accordingly, and comparable convergence trends are observed in limited number of shots. These observations support our theoretical analysis, which shows that even with a smaller scale factor on the penalty, the sampling overhead associated with the penalty term does not increase significantly.

\section{Conclusions}\label{sec:conclusions}
In this work, we presented a spectral machine learning framework implemented with compact quantum circuit designs.
The proposed quantum circuits realize spectral clustering while avoiding two common bottlenecks of approximation-based pipelines: the explicit construction of the full or partial kernel matrix, and the separate estimation of individual quantum fidelity kernel entries.
In particular, we rigorously showed that the shot requirement for the quantity estimated by the penalty circuit is independent of the training dataset size.
This mitigates the apparent scale mismatch between the penalty and the other terms in the objective function.

As a proof of concept, we validated the proposed method through numerical simulations, showing that it can reliably learn the structure of the training data and assign cluster membership to unseen samples using a parameterized quantum model.
We also observed empirically that estimating the penalty during training typically does not require a large number of shots: the penalty estimates exhibited similar trends across shot numbers of 256, 512, 1024, and 8192, consistent with the theorem presented in the main text.

In the preceding section, we discussed how our approach is distinguished by its ability to efficiently train objectives that rely on the quantum fidelity kernel. Whether the quantum fidelity kernel itself is broadly useful for general machine learning tasks, however, remains an open question. Since this work focuses on designing more efficient quantum circuits for the case in which quantum kernel methods are used as the primary tool for unsupervised learning, we do not directly address the quantum advantage conferred by the encoding-induced kernel itself. Nevertheless, as the study of quantum kernel methods continues to mature, the demand for more efficient implementations is likely to grow, and we expect our methodology to contribute meaningfully in that direction.

Several avenues for further research remain. First, quantum kernel based methods have been studied extensively, and many of their properties are well understood. However, most of this literature concerns supervised learning, which differs substantially from our focus on unsupervised learning—where performance depends heavily on the geometry of the feature-mapped data. In this work, we fix the feature map via a prescribed encoding circuit, thereby maintaining a consistent geometry of the induced distribution in Hilbert space. To our knowledge, no prior study identifies advantageous regimes for quantum encodings in the unsupervised setting.

In parallel, spectral graph theory offers a rich toolkit ready for rigorous use. Unlike the standard kernel matrix (treated here as an `adjacency matrix'), the graph Laplacian has distinctive properties and a mature theory of its own. Although this paper focused on efficient circuit implementations for Laplacian-based machine-learning tasks, in the future work we will leverage these tools to develop a robust theoretical account of quantum machine learning pipelines for graph-based data representations. We believe that, despite differences in learning objectives, analytical frameworks developed for supervised quantum kernels can be adapted to the unsupervised setting when combined with techniques from spectral graph theory.

Second, we rigorously showed that the quantity estimated by the penalty circuit can be estimated reliably without a significant increase in the overall shot budget.
This raises the natural question of whether the same approach can be extended to other variational quantum algorithms with analogous objective structures. 
For instance, many combinatorial optimization tasks are formulated as quadratic unconstrained binary optimization (QUBO) problems and solved using variational quantum circuits whose objectives include one or more penalty terms. 
Because such formulations often involve multiple penalties, our analysis may help reduce the overall query (shot) requirements in practical implementations on real quantum hardware.

Finally, our analysis of the penalty estimator leverages concentration of measure in the exponentially high-dimensional Hilbert space. 
Notably, the same concentration phenomenon is often implicated in well-known obstacles to the trainability of variational quantum algorithms \cite{BarrenPlateausMcclean2018}.
In our setting, by contrast, it acts in the opposite direction, helping to explain why the shot complexity required to estimate the penalty term can remain comparatively small.
More comprehensive understandings of these dual roles as both an obstacle to trainability and an aid to sampling are interesting directions for future works.

\medskip
\textbf{Code Availability} \par
The code used for data processing and analysis is available at GitHub: \href{https://github.com/hyeong-gyu-kim/qsc}{https://github.com/hyeong-gyu-kim/qsc}.

\medskip
\textbf{Acknowledgements} \par 
This work was supported by the National Research Foundation of Korea(NRF) grant funded by the Korea government(MSIT) (No.RS-2021-NR059555, No.RS-2026-25482650), and by Qunova Computing, Inc. (Development of Algorithms for Quantum Machine Learning and Discovery of Application Cases).

\medskip

\bibliographystyle{unsrt}
\bibliography{bibliography}

@book{golub2013matrix,
  title={Matrix computations},
  author={Golub, Gene H and Van Loan, Charles F},
  year={2013},
  publisher={JHU press}
}

@article{UniformlyControlledGate2,
  title={Quantum circuits with uniformly controlled one-qubit gates},
  author={Bergholm, Ville and Vartiainen, Juha J and M{\"o}tt{\"o}nen, Mikko and Salomaa, Martti M},
  journal={Physical Review A—Atomic, Molecular, and Optical Physics},
  volume={71},
  number={5},
  pages={052330},
  year={2005},
  publisher={APS}
}

@article{Lanczos,
  title={An iteration method for the solution of the eigenvalue problem of linear differential and integral operators},
  author={Lanczos, Cornelius},
  journal={Journal of research of the National Bureau of Standards},
  volume={45},
  number={4},
  pages={255--282},
  year={1950}
}

@article{RFF,
  title={Random features for large-scale kernel machines},
  author={Rahimi, Ali and Recht, Benjamin},
  journal={Advances in neural information processing systems},
  volume={20},
  year={2007}
}

@article{Nystrom,
  title={Using the Nystr{\"o}m method to speed up kernel machines},
  author={Williams, Christopher and Seeger, Matthias},
  journal={Advances in neural information processing systems},
  volume={13},
  year={2000}
}

@inproceedings{ComplexProj2designs,
  title={Mutually unbiased bases are complex projective 2-designs},
  author={Klappenecker, Andreas and Rotteler, Martin},
  booktitle={Proceedings. International Symposium on Information Theory, 2005. ISIT 2005.},
  pages={1740--1744},
  year={2005},
  organization={IEEE}
}

@book{HighDimensionalProbability,
  title={High-dimensional probability: An introduction with applications in data science},
  author={Vershynin, Roman},
  volume={47},
  year={2018},
  publisher={Cambridge university press}
}

@inproceedings{RayleighConvergence,
  title={Tight convergence rate of gradient descent for eigenvalue computation},
  author={Ding, Qinghua and Zhou, Kaiwen and Cheng, James},
  booktitle={Proceedings of the Twenty-Ninth International Conference on International Joint Conferences on Artificial Intelligence},
  pages={3276--3282},
  year={2021}
}

@article{ParamShift1,
  title={Quantum circuit learning},
  author={Mitarai, Kosuke and Negoro, Makoto and Kitagawa, Masahiro and Fujii, Keisuke},
  journal={Physical Review A},
  volume={98},
  number={3},
  pages={032309},
  year={2018},
  publisher={APS}
}

@article{ParamShift2,
  title={Evaluating analytic gradients on quantum hardware},
  author={Schuld, Maria and Bergholm, Ville and Gogolin, Christian and Izaac, Josh and Killoran, Nathan},
  journal={Physical Review A},
  volume={99},
  number={3},
  pages={032331},
  year={2019},
  publisher={APS}
}

@inproceedings{QuantumFanout,
  title={Quantum fan-out: Circuit optimizations and technology modeling},
  author={Gokhale, Pranav and Koretsky, Samantha and Huang, Shilin and Majumder, Swarnadeep and Drucker, Andrew and Brown, Kenneth R and Chong, Frederic T},
  booktitle={2021 IEEE International Conference on Quantum Computing and Engineering (QCE)},
  pages={276--290},
  year={2021},
  organization={IEEE}
}

@article{KFuji,
  title={Quantum circuit learning},
  author={Mitarai, Kosuke and Negoro, Makoto and Kitagawa, Masahiro and Fujii, Keisuke},
  journal={Physical Review A},
  volume={98},
  number={3},
  pages={032309},
  year={2018},
  publisher={APS}
}

@article{VDunjko,
  title={Quantum-enhanced machine learning},
  author={Dunjko, Vedran and Taylor, Jacob M and Briegel, Hans J},
  journal={Physical review letters},
  volume={117},
  number={13},
  pages={130501},
  year={2016},
  publisher={APS}
}

@article{JacobB,
  title={Quantum machine learning},
  author={Biamonte, Jacob and Wittek, Peter and Pancotti, Nicola and Rebentrost, Patrick and Wiebe, Nathan and Lloyd, Seth},
  journal={Nature},
  volume={549},
  number={7671},
  pages={195--202},
  year={2017},
  publisher={Nature Publishing Group UK London}
}

@article{MBenedetti,
  title={Parameterized quantum circuits as machine learning models},
  author={Benedetti, Marcello and Lloyd, Erika and Sack, Stefan and Fiorentini, Mattia},
  journal={Quantum science and technology},
  volume={4},
  number={4},
  pages={043001},
  year={2019},
  publisher={IOP Publishing}
}

@article{ICong,
  title={Quantum convolutional neural networks},
  author={Cong, Iris and Choi, Soonwon and Lukin, Mikhail D},
  journal={Nature Physics},
  volume={15},
  number={12},
  pages={1273--1278},
  year={2019},
  publisher={Nature Publishing Group UK London}
}

@article{SMangini,
  title={Quantum computing models for artificial neural networks},
  author={Mangini, Stefano and Tacchino, Francesco and Gerace, Dario and Bajoni, Daniele and Macchiavello, Chiara},
  journal={Europhysics Letters},
  volume={134},
  number={1},
  pages={10002},
  year={2021},
  publisher={IOP Publishing}
}

@article{PatrickR,
  title={Quantum support vector machine for big data classification},
  author={Rebentrost, Patrick and Mohseni, Masoud and Lloyd, Seth},
  journal={Physical review letters},
  volume={113},
  number={13},
  pages={130503},
  year={2014},
  publisher={APS}
}

@article{Mschuld,
  title={Implementing a distance-based classifier with a quantum interference circuit},
  author={Schuld, Maria and Fingerhuth, Mark and Petruccione, Francesco},
  journal={Europhysics Letters},
  volume={119},
  number={6},
  pages={60002},
  year={2017},
  publisher={IOP Publishing}
}

@article{DKPark,
  title={The theory of the quantum kernel-based binary classifier},
  author={Park, Daniel K and Blank, Carsten and Petruccione, Francesco},
  journal={Physics Letters A},
  volume={384},
  number={21},
  pages={126422},
  year={2020},
  publisher={Elsevier}
}

@article{CBlank,
  title={Quantum classifier with tailored quantum kernel},
  author={Blank, Carsten and Park, Daniel K and Rhee, June-Koo Kevin and Petruccione, Francesco},
  journal={npj Quantum Information},
  volume={6},
  number={1},
  pages={41},
  year={2020},
  publisher={Nature Publishing Group UK London}
}

@article{SPark,
  title={Variational quantum approximate support vector machine with inference transfer},
  author={Park, Siheon and Park, Daniel K and Rhee, June-Koo Kevin},
  journal={Scientific reports},
  volume={13},
  number={1},
  pages={3288},
  year={2023},
  publisher={Nature Publishing Group UK London}
}

@article{QPCA,
  title={Quantum principal component analysis},
  author={Lloyd, Seth and Mohseni, Masoud and Rebentrost, Patrick},
  journal={Nature physics},
  volume={10},
  number={9},
  pages={631--633},
  year={2014},
  publisher={Nature Publishing Group UK London}
}

@article{HHL,
  title={Quantum algorithm for linear systems of equations},
  author={Harrow, Aram W and Hassidim, Avinatan and Lloyd, Seth},
  journal={Physical review letters},
  volume={103},
  number={15},
  pages={150502},
  year={2009},
  publisher={APS}
}

@article{Scikit,
  title={Scikit-learn: Machine learning in Python},
  author={Pedregosa, Fabian and Varoquaux, Ga{\"e}l and Gramfort, Alexandre and Michel, Vincent and Thirion, Bertrand and Grisel, Olivier and Blondel, Mathieu and Prettenhofer, Peter and Weiss, Ron and Dubourg, Vincent and others},
  journal={the Journal of machine Learning research},
  volume={12},
  pages={2825--2830},
  year={2011},
  publisher={JMLR. org}
}

@article{Pennylane,
  title={Pennylane: Automatic differentiation of hybrid quantum-classical computations},
  author={Bergholm, Ville and Izaac, Josh and Schuld, Maria and Gogolin, Christian and Ahmed, Shahnawaz and Ajith, Vishnu and Alam, M Sohaib and Alonso-Linaje, Guillermo and AkashNarayanan, B and Asadi, Ali and others},
  journal={arXiv preprint arXiv:1811.04968},
  year={2018}
}

@article{SukinS,
  title={Expressibility and entangling capability of parameterized quantum circuits for hybrid quantum-classical algorithms},
  author={Sim, Sukin and Johnson, Peter D and Aspuru-Guzik, Al{\'a}n},
  journal={Advanced Quantum Technologies},
  volume={2},
  number={12},
  pages={1900070},
  year={2019},
  publisher={Wiley Online Library}
}

@article{Suyken_SVMPCA,
author = {Suykens, Johan and Van Gestel, Tony and Vandewalle, Joos and De Moor, Bart},
year = {2003},
month = {04},
pages = {447 - 450},
title = {A Support Vector Machine Formulation to PCA Analysis and its Kernel Version},
volume = {14},
journal = {Neural Networks, IEEE Transactions on},
doi = {10.1109/TNN.2003.809414}
}

@INPROCEEDINGS{Suykens_WPCA,
  author={Alzate, C. and Suykens, J.A.K.},
  booktitle={The 2006 IEEE International Joint Conference on Neural Network Proceedings}, 
  title={A Weighted Kernel PCA Formulation with Out-of-Sample Extensions for Spectral Clustering Methods}, 
  year={2006},
  volume={},
  number={},
  pages={138-144},
  doi={10.1109/IJCNN.2006.246671}}

@misc{TutorialSC,
      title={A Tutorial on Spectral Clustering}, 
      author={Ulrike von Luxburg},
      year={2007},
      eprint={0711.0189},
      archivePrefix={arXiv},
      primaryClass={cs.DS},
      url={https://arxiv.org/abs/0711.0189}, 
}

@article{RandIndex,
  title={Objective criteria for the evaluation of clustering methods},
  author={Rand, William M},
  journal={Journal of the American Statistical association},
  volume={66},
  number={336},
  pages={846--850},
  year={1971},
  publisher={Taylor \& Francis}
}

@article{ARI,
  title={Comparing partitions},
  author={Hubert, Lawrence and Arabie, Phipps},
  journal={Journal of classification},
  volume={2},
  pages={193--218},
  year={1985},
  publisher={Springer}
}

@article{UniformlyControlledGate,
  title={Transformation of quantum states using uniformly controlled rotations},
  author={Mottonen, Mikko and Vartiainen, Juha J and Bergholm, Ville and Salomaa, Martti M},
  journal={arXiv preprint quant-ph/0407010},
  year={2004}
}

@book{SpectralGraphChung1997,
  title={Spectral graph theory},
  author={Chung, Fan RK},
  volume={92},
  year={1997},
  publisher={American Mathematical Soc.}
}

@article{SemiSupervisedKipf2016,
  title={Semi-supervised classification with graph convolutional networks},
  author={Kipf, Thomas N and Welling, Max},
  journal={arXiv preprint arXiv:1609.02907},
  year={2016}
}

@article{OnSpectralClusteringNg2001,
  title={On spectral clustering: Analysis and an algorithm},
  author={Ng, Andrew and Jordan, Michael and Weiss, Yair},
  journal={Advances in neural information processing systems},
  volume={14},
  year={2001}
}

@article{NormalizedCutsShi2000,
  title={Normalized cuts and image segmentation},
  author={Shi, Jianbo and Malik, Jitendra},
  journal={IEEE Transactions on pattern analysis and machine intelligence},
  volume={22},
  number={8},
  pages={888--905},
  year={2000},
  publisher={Ieee}
}

@article{RandomWalksMelia2000,
  title={Learning segmentation by random walks},
  author={Meila, Marina and Shi, Jianbo},
  journal={Advances in neural information processing systems},
  volume={13},
  year={2000}
}

@inproceedings{SegmentationWeiss1999,
  title={Segmentation using eigenvectors: a unifying view},
  author={Weiss, Yair},
  booktitle={Proceedings of the seventh IEEE international conference on computer vision},
  volume={2},
  pages={975--982},
  year={1999},
  organization={IEEE}
}

@article{LaplacianEigenmapsBelkin2003,
  title={Laplacian eigenmaps for dimensionality reduction and data representation},
  author={Belkin, Mikhail and Niyogi, Partha},
  journal={Neural computation},
  volume={15},
  number={6},
  pages={1373--1396},
  year={2003},
  publisher={MIT Press}
}

@article{QuantumKernelSchuld2019,
  title={Quantum machine learning in feature Hilbert spaces},
  author={Schuld, Maria and Killoran, Nathan},
  journal={Physical review letters},
  volume={122},
  number={4},
  pages={040504},
  year={2019},
  publisher={APS}
}

@article{QuantumKernelRealDeviceHavlivcek2019,
  title={Supervised learning with quantum-enhanced feature spaces},
  author={Havl{\'\i}{\v{c}}ek, Vojt{\v{e}}ch and C{\'o}rcoles, Antonio D and Temme, Kristan and Harrow, Aram W and Kandala, Abhinav and Chow, Jerry M and Gambetta, Jay M},
  journal={Nature},
  volume={567},
  number={7747},
  pages={209--212},
  year={2019},
  publisher={Nature Publishing Group UK London}
}

@article{BarrenPlateausMcclean2018,
  title={Barren plateaus in quantum neural network training landscapes},
  author={McClean, Jarrod R and Boixo, Sergio and Smelyanskiy, Vadim N and Babbush, Ryan and Neven, Hartmut},
  journal={Nature communications},
  volume={9},
  number={1},
  pages={4812},
  year={2018},
  publisher={Nature Publishing Group UK London}
}

@article{PCAHotelling1933,
  title={Analysis of a complex of statistical variables into principal components.},
  author={Hotelling, Harold},
  journal={Journal of educational psychology},
  volume={24},
  number={6},
  pages={417},
  year={1933},
  publisher={Warwick \& York}
}

@article{PCAPearson1901,
  title={LIII. On lines and planes of closest fit to systems of points in space},
  author={Pearson, Karl},
  journal={The London, Edinburgh, and Dublin philosophical magazine and journal of science},
  volume={2},
  number={11},
  pages={559--572},
  year={1901},
  publisher={Taylor \& Francis}
}

@book{OptimizationOnMatrixManifolds,
  title={Optimization algorithms on matrix manifolds},
  author={Absil, P-A and Mahony, Robert and Sepulchre, Rodolphe},
  year={2008},
  publisher={Princeton University Press}
}

\newpage
\appendix

\begin{figure}[t]
    \centering
    \begin{minipage}[t]{0.8\textwidth}
        \centering
        \includegraphics[width=\textwidth]{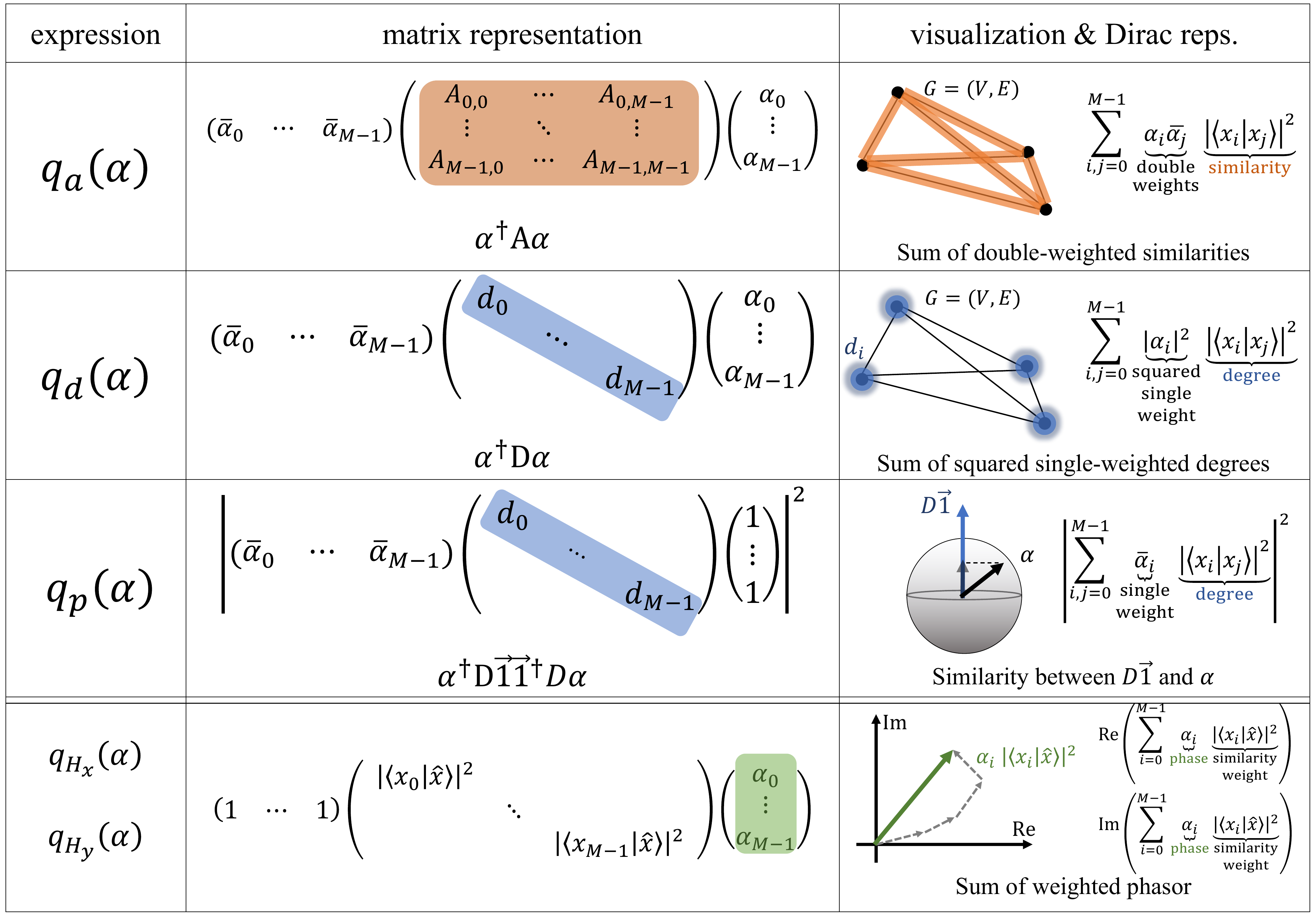}
    \end{minipage}%
    \hspace{0.1cm} 
    \caption{Alternative visualizations of the quantum circuits shown in Figure \ref{fig:circuits}. Here, $d_i := (\bm{D})_{ii}=\sum_{j=0}^{M-1} |\langle \bm{x_i} | \bm{x_j}\rangle|^2$.}
    \label{fig:Visualization}
\end{figure}

\section{Estimator circuits}\label{app:estimators}
This appendix collects derivations of $q_a$, $q_d$, and $q_p$ and states the exact scaling used to compute $\mathcal{J}(\bm{\alpha})$.
The operational interpretation of each estimator circuit is illustrated in \textbf{Figure \ref{fig:Visualization}}.

\subsection{Adjacency-form estimator $q_a$}
Here, we introduce an efficient quantum circuit architecture for computing the quadratic form of the graph Laplacian matrix corresponds to $\bm{\alpha^{\dagger} A \alpha}$ constructed from the given dataset $\mathcal{D}$. For a visual summary, see \textbf{Figure \ref{fig:circuits-a}}, which illustrates the circuit used in the following discussion.

\begin{enumerate}
    \item Initialize the quantum state as
    \begin{align}\label{eqn:adj_start}
        |\psi_{\text{a,1}}\rangle = \sum_{i=0}^{M-1} \alpha_i |i\rangle_{A_1} |0\rangle_{B_1} |i\rangle_{A_2} |0\rangle_{B_2} \in (\mathbb{C}^2)^{\otimes m}\otimes(\mathbb{C}^2)^{\otimes n}\otimes(\mathbb{C}^2)^{\otimes m}\otimes(\mathbb{C}^2)^{\otimes n}
    \end{align} 
    by applying the weighting unitary $U_\alpha$ and CNOT gates to the pairs of index registers.
    \item Apply \(U_{\varphi, \mathcal{D}}\) and its complex conjugate \(\overline{U}_{\varphi, \mathcal{D}}\) to obtain 
    \begin{align}
        |\psi_{\text{a,2}}\rangle = (U_{\varphi, \mathcal{D}} \otimes \overline{U}_{\varphi, \mathcal{D}})|\psi_{\text{a,1}}\rangle = \sum_{i=0}^{M-1} \alpha_i |i\rangle_{A_1} |\bm{x_i}\rangle_{B_1} |i\rangle_{A_2} |\bm{\overline{x}_i}\rangle_{B_2},
    \end{align}
    where \(|\bm{\overline{x}}\rangle\) denotes the the state prepared by \(\overline{U}_{\varphi, \mathcal{D}}\).
    \item Untangle the index registers by applying CNOT gates:
    \begin{align}
        |\psi_{\text{a,3}}\rangle = \sum_{i=0}^{M-1} \alpha_i |i\rangle_{A_1} |\bm{x_i}\rangle_{B_1} |0\rangle_{A_2} |\bm{\overline{x}_i}\rangle_{B_2}.
    \end{align}
    \item Apply Hadamard gates to the first index register: 
    \begin{align}
        |\psi_{\text{a,4}}\rangle = H^{\otimes m}_{A_1} |\psi_{\text{a,3}}\rangle.
    \end{align}
    \item Perform the projective measurement $\mathcal{M}_a := |0^m \rangle\langle 0^m|_{A_1} \otimes I_{B_1} \otimes I_{A_2} \otimes I_{B_2}$, which yields 
    \begin{align}\label{eqn:adj_end}
        q_a := \operatorname{Tr}\left(|\psi_{a,4}\rangle\langle \psi_{a,4}| \mathcal{M}_a\right) = M^{-1} \sum_{i,j=0}^{M-1} \alpha_i \overline{\alpha}_j |\langle \bm{x_i} | \bm{x_j} \rangle|^2.
    \end{align}
\end{enumerate}

One may observe that this is precisely equivalent the quadratic form of the kernel matrix built from the fidelity kernel, $\kappa(\bm{x_i}, \bm{x_j}) := |\langle \bm{x_i} | \bm{x_j} \rangle|^2$, a standard choice in quantum machine learning. This observation indicates that the proposed circuit applies not only to graph Laplacian analysis but also to other quantum kernel-based learning tasks, such as quantum support vector machines.

\subsection{Degree-form estimator $q_d$}
The computation of the degree-weighted quadratic form $\alpha^\dagger D \alpha$ proceeds analogously but incorporates a SWAP-test. For a visual summary, see \textbf{Figure \ref{fig:circuits-b}}, which illustrates the circuit used in the discussion below.

\begin{enumerate}
    \item Initialize the quantum state as 
    \begin{align}\label{eqn:deg_start}
        |\psi_{\text{d,1}}\rangle &=M^{-1/2} \sum_{i,j=0}^{M-1}  \alpha_i |i\rangle_{A_1} |0\rangle_{B_1} |j\rangle_{A_2} |0\rangle_{B_2} |0\rangle_{C} \\ &\in (\mathbb{C}^2)^{\otimes m}\otimes(\mathbb{C}^2)^{\otimes n}\otimes(\mathbb{C}^2)^{\otimes m}\otimes(\mathbb{C}^2)^{\otimes n}\otimes \mathbb{C}^2 \nonumber
    \end{align} 
    This is prepared by applying the weighting unitary $U_\alpha$ on $A_1$, Hadamard gates on $A_2$, and leaving the ancilla $C$ in $|0\rangle$.
    \item Apply $U_{\varphi, \mathcal{D}}$ to the data registers $B_1$ and $B_2$:
    \begin{align}
        |\psi_{\text{d,2}}\rangle &= \left((U_{\varphi, \mathcal{D}})_{A_1, B_1} \otimes (U_{\varphi, \mathcal{D}})_{A_2, B_2} \otimes I_C \right)|\psi_{\text{d,1}}\rangle \\
        &=M^{-1/2} \sum_{i,j=0}^{M-1} \alpha_i |i\rangle_{A_1} |\bm{x_i}\rangle_{B_1} |j\rangle_{A_2} |\bm{x_j}\rangle_{B_2} |0\rangle_{C}. \nonumber
    \end{align}
    \item Apply $\textit{SWAP-Test}_{(W,V|S)} = H_S \cdot \textit{C-SWAP}_{(W,V|S)} \cdot H_S$, where $\textit{C-SWAP}_{(W,V|S)}$ denotes the controlled-SWAP operator that swaps registers $W$ and $V$ conditioned on the ancilla qubit $S$:
    \begin{align}
        |\psi_{\text{d,3}}\rangle &= \textit{SWAP-Test}_{(B_1, B_2|C)}|\psi_{\text{d,2}}\rangle \\
        &= (4M)^{-1/2} \sum_{i,j=0}^{M-1} \alpha_i \Big(|i\rangle_{A_1} |\bm{x_i}\rangle_{B_1} |j\rangle_{A_2} |\bm{x_j}\rangle_{B_2} + |i\rangle_{A_1} |\bm{x_j}\rangle_{B_1} |j\rangle_{A_2} |\bm{x_i}\rangle_{B_2} \Big) |0\rangle_{C} \nonumber \\
        &+ \alpha_i \Big( |i\rangle_{A_1} |\bm{x_i}\rangle_{B_1} |j\rangle_{A_2} |\bm{x_j}\rangle_{B_2} - |i\rangle_{A_1} |\bm{x_j}\rangle_{B_1} |j\rangle_{A_2} |\bm{x_i}\rangle_{B_2} \Big) |1\rangle_{C}. \nonumber
    \end{align}
    \item Perform the projective measurement $\mathcal{M}_d := I_{A_1} \otimes I_{B_1} \otimes I_{A_2} \otimes I_{B_2} \otimes (\sigma_z)_C$, which yields 
    \begin{align}\label{eqn:deg_end_appendix}
        q_d := \operatorname{Tr}(|\psi_{d,3}\rangle \langle \psi_{d,3}| \mathcal{M}_d) = M^{-1} \sum_{i,j=0}^{M-1} |\alpha_i|^2 |\langle \bm{x_i} | \bm{x_j} \rangle|^2.
    \end{align}
\end{enumerate}

Since $(\bm{D})_{ii}=\sum_{j=0}^{M-1} |\langle \bm{x_i} | \bm{x_j} \rangle|^2$, it is straightforward to see that Equation \ref{eqn:deg_end_appendix} represents the summation over the diagonal elements of $\bm{D}$, weighted by the squared entries of an arbitrary vector $\bm{\alpha}$.

\subsection{Projector penalty estimator $q_p$}
Computation of the quadratic form of the $D \vec{1} \vec{1}^{\dagger} D$ projector term is more delicate, owing to both the required classical post-processing and the scale mismatch with the previous quadratic-form terms. Unlike the earlier cases, this step requires additional subroutine to evaluate $\operatorname{Tr}(|+\rangle \langle +|^{\otimes m} \alpha \alpha^\dagger)$. For a visual summary, see \textbf{Figure \ref{fig:circuits-c}}, which illustrates the corresponding circuit.

\begin{enumerate}
    \item Initialize the quantum state as 
    \begin{align}\label{eqn:deg_start}
        |\psi_{\text{p,1}}\rangle &= M^{-1/2} \sum_{i,j=0}^{M-1}  \alpha_i  |i\rangle_{A_0} |j\rangle_{A_1} |0\rangle_{B_1} |j\rangle_{A_2} |0\rangle_{B_2} |0\rangle_{C} \\ &\in (\mathbb{C}^2)^{\otimes m}\otimes(\mathbb{C}^2)^{\otimes m}\otimes(\mathbb{C}^2)^{\otimes n}\otimes(\mathbb{C}^2)^{\otimes m}\otimes(\mathbb{C}^2)^{\otimes n} \otimes \mathbb{C}^2. \nonumber
    \end{align} 
    This is prepared by applying the weighting unitary $U_\alpha$ on $A_0$, Hadamard gates $H^{\otimes m}$ on $A_1$, CNOTs from $A_1$ to $A_2$, and keeping the ancilla $C$ in $|0\rangle$.
    \item Apply \(U_{\varphi, \mathcal{D}}\) to the data registers \(B_1\) and \(B_2\):
    \begin{align}
        |\psi_{\text{p,2}}\rangle &= M^{-1/2} \left(I_{A_0} \otimes (U_{\varphi, \mathcal{D}})_{A_1, B_1} \otimes (\overline{U}_{\varphi, \mathcal{D}})_{A_2, B_2} \otimes I_C \right)|\psi_{\text{p,1}}\rangle \\
        &=\sum_{i,j=0}^{M-1} \alpha_i |i\rangle_{A_0} |j\rangle_{A_1} |\bm{x_j}\rangle_{B_1} |j\rangle_{A_2} |\bm{\overline{x}_j}\rangle_{B_2} |0\rangle_{C}. \nonumber
    \end{align}
    \item Untangle $A_1$ and $A_2$ with CNOTs:
    \begin{align}
        |\psi_{\text{p,3}}\rangle = M^{-1/2} \sum_{i,j=0}^{M-1} \alpha_i |i\rangle_{A_0} |j\rangle_{A_1} |\bm{x_j}\rangle_{B_1} |0\rangle_{A_2} |\bm{\overline{x}_j}\rangle_{B_2} |0\rangle_{C}.
    \end{align}
    \item Apply a Hadamard on the ancilla $C$ and a controlled-SWAP between $A_0$ and $A_1$ with control $C$:
    \begin{align}
        |\psi_{\text{p,4}}\rangle &= \textit{C-SWAP}_{(A_0, A_1|C)}|\psi_{\text{p,3}}\rangle \\
        &= (2M)^{-1/2} \sum_{i,j=0}^{M-1} \alpha_i \Big( |i\rangle_{A_0} |j\rangle_{A_1} |\bm{x_j}\rangle_{B_1} |0\rangle_{A_2} |\bm{\overline{x}_j}\rangle_{B_2} |0\rangle_{C} + |j\rangle_{A_0} |i\rangle_{A_1} |\bm{x_j}\rangle_{B_1} |0\rangle_{A_2} |\bm{\overline{x}_j}\rangle_{B_2} |1\rangle_{C} \Big)
    \end{align}
    \item Perform the projective measurements $\mathcal{M}_{p,x} := |+ \rangle\langle +|^{\otimes m}_{A_0} \otimes I_{A_1} \otimes I_{B_1} \otimes I_{A_2} \otimes I_{B_2} \otimes (\sigma_x)_C$ and $\mathcal{M}_{p,y} := |+ \rangle\langle +|^{\otimes m}_{A_0} \otimes I_{A_1} \otimes I_{B_1} \otimes I_{A_2} \otimes I_{B_2} \otimes (\sigma_y)_C$, which yields 
    \begin{align}\label{eqn:deg_end}
        p_x := \operatorname{Tr}(|\psi_{p,4}\rangle \langle \psi_{p,4}| \mathcal{M}_{p,x}) = \operatorname{Re }M^{-2} \left(\sum_{t=0}^{M-1} \alpha_t \right) \left(\sum_{i,k=0}^{M-1} \overline{\alpha}_i |\langle \bm{x_i} | \bm{x_k} \rangle|^2\right), \\
        p_y := \operatorname{Tr}(|\psi_{p,4}\rangle \langle \psi_{p,4}| \mathcal{M}_{p,y}) = \operatorname{Im }M^{-2} \left(\sum_{t=0}^{M-1} \alpha_t \right) \left(\sum_{i,k=0}^{M-1} \overline{\alpha}_i |\langle \bm{x_i} | \bm{x_k} \rangle|^2\right),
    \end{align}
    respectively.
    \item Compute $p_x^2 + p_y^2$ and divide by $\left|M^{-1}\sum_{t=0}^{M-1} \alpha_t \right|^2$, which is obtained from an independent circuit that estimates $\operatorname{Tr}(|+\rangle \langle+|^{\otimes m} \alpha \alpha^\dagger)$. The final expression is
    \begin{align}{\label{eqn:proj_end}}
        q_p(\alpha) := M^{-2} \left|\sum_{i,k=0}^{M-1} \overline{\alpha}_i  |\langle \bm{x_i} | \bm{x_k} \rangle|^2\right|^2.
    \end{align}
\end{enumerate}

\section{Score estimator circuit}\label{app:test-circuits}
In this section, we briefly explain the computation of $q_{H_x}$ and $q_{H_y}$ using the circuits shown in \textbf{Figure \ref{fig:circuits-d}}. It computes $q_{H_x}$ and $q_{H_y}$ through the following steps. For the prepared state 
\begin{align}
    |\psi_{H,1}\rangle = (M)^{-1/2}\sum_{i=0}^{M-1} |0\rangle_{C_1} |i\rangle_{A_1} |\bm{x_i}\rangle_{B_1} |\bm{\hat{x}}\rangle_{B_2} |0\rangle_{C_2}
\end{align} 
generated via the unitary operation $U_{\varphi,\mathcal{D}}$, we apply $\textit{Hadamard-Test}_{(A_1|C_1)}:= |0\rangle \langle 0|_{C_1} \otimes (H^{\otimes m})_{A_1} + |1\rangle\langle 1|_{C_1} \otimes (U_{\alpha})_{A_1}$ and $\textit{SWAP-Test}_{(B_1,B_2|C_2)}$. As before, the SWAP test computes the fidelity between the encoded training data and the test data. The Hadamard test is employed to examine the real and imaginary parts of the sum of local phase in the index register $A_1$, which can be retrieved by measuring register $C_1$ in the $\sigma_x$ and $\sigma_y$ bases, respectively. This becomes clear by examining the quantum state after measuring $C_1$ and $C_2$ with the $(\sigma_x$, $\sigma_y)$ and $\sigma_z$ respectively,
\begin{align}
    &\frac{1}{2\sqrt{M}} \sum_{i,j=0}^{M-1} \left( \alpha_i + \alpha_j^* \right) \big( |i\rangle\langle j| \otimes |\bm{\hat{x}}\rangle\langle \bm{x_j}| \otimes |\bm{x_i}\rangle\langle \bm{\hat{x}}| + |i\rangle\langle j| \otimes |\bm{x_i}\rangle \langle \bm{\hat{x}}| \otimes |\bm{\hat{x}}\rangle \langle \bm{x_j}| \big)_{(A_1, B_1, B_2)}, \\
    &\frac{1}{2\sqrt{M}} \sum_{i,j=0}^{M-1} \left( \alpha_i - \alpha_j^* \right) \big( |i\rangle\langle j| \otimes |\bm{\hat{x}}\rangle\langle \bm{x_j}| \otimes |\bm{x_i}\rangle\langle \bm{\hat{x}}| + |i\rangle\langle j| \otimes |\bm{x_i}\rangle \langle \bm{\hat{x}}| \otimes |\bm{\hat{x}}\rangle \langle \bm{x_j}| \big)_{(A_1, B_1, B_2)}.
\end{align}
Tracing out the all remaining qubits yield
\begin{align}
    \label{eqn:score_end1}
    &\frac{1}{2\sqrt{M}} \sum_i \left( \alpha_i + \alpha_i^* \right) \cdot |\langle \bm{\hat{x}} \mid \bm{x_i} \rangle|^2 = \frac{1}{\sqrt{M}} \sum_i \operatorname{Re}(\alpha_i) |\langle \bm{\hat{x}} \mid \bm{x_i} \rangle|^2, \\
    \label{eqn:score_end2}
    &\frac{1}{2\sqrt{M}} \sum_i  \left( \alpha_i - \alpha_i^* \right) \cdot |\langle \bm{\hat{x}} \mid \bm{x_i} \rangle|^2 = \frac{1}{\sqrt{M}} \sum_i \operatorname{Im}(\alpha_i) |\langle \bm{\hat{x}} \mid \bm{x_i} \rangle|^2.
\end{align}
See supplementary for the proof with explained in step-by-step.

Although the circuit shown in \textbf{Figure \ref{fig:circuits-d}} may appear to have nontrivial depth due to the \textit{Hadamard-test} routine, its overall depth scaling remains comparable to that of the other estimator circuits, since one-control, multiple-target quantum gates can be implemented with only a constant number of additional gates while avoiding the serialization of single-qubit gates \cite{QuantumFanout}, so that the depth of controlled-$U_\alpha$ is increased only by a constant factor.

\section{Proof of Theorem 1}\label{app:shot-proof-sketch}
A potential concern is the scale mismatch between Equation \ref{eqn:adj_end} and \ref{eqn:deg_end_appendix}, which yield quantities of order $1 / M$, and Equation \ref{eqn:proj_end}, which yields values of order $1 / M^2$. The sampling requirement associated with Equation \ref{eqn:proj_end} can be analyzed rigorously via concentration inequalities. In our formulation, Equation \ref{eqn:proj_end} can be viewed as defining a loss function evaluated at a random variable $\bm{\alpha} \sim \mu$, where $\mu$ is a distribution of our choice. To facilitate the analysis, we introduce the auxiliary quantity $Z(\bm{\alpha})$ that isolates the penalty term $q_p(\bm{\alpha})$ from the objective:
\begin{align}\label{eqn:penalty_Z}
Z(\bm{\alpha}):= \frac{1}{M^{2}}|\langle \bm{d}, \bm{\alpha}\rangle||\langle \bm{\vec{1}}, \bm{\alpha}\rangle|,    
\end{align}
where $\bm{d}:=\left(d_0, \ldots, d_{M-1}\right)^{\top} \in \mathbb{R}^M$ and $d_i:=\sum_{j=0}^{M-1}\left|\left\langle x_i, x_j\right\rangle\right|^2$. Under the stable convergence of Equation \ref{eq:objective-main} (see Supplementary Material for the convergence of the Rayleigh quotient), and the notation $t$ corresponds to an iteration step during the optimization, we introduce the following lemmas. 

\begin{lemma}
Suppose $Z(\bm{\alpha}^{(t)}) \xrightarrow{\mathbb{P}} 0$ as $t\to \infty$ and, for some constant $B<\infty$, we have $0\le Z(\bm{\alpha}^{(t)})\le B$ almost surely for all $t$. Then
$$
\mathbb{E}[Z(\bm{\alpha}^{(t)})] \longrightarrow 0,\qquad \mathrm{Var}[Z(\bm{\alpha}^{(t)})] \longrightarrow 0 \qquad (t\to \infty).
$$
\end{lemma}

\begin{lemma}
Let $M\ge 1$, let $\bm{\alpha}\in\mathbb{C}^M$ be Haar--uniform on the unit sphere,
and let $\bm{d} \in \mathbb{R}^M$. Then
$$
\mathbb{E}[Z] \le \frac{1}{M}, \qquad \mathrm{Var}[Z] \le \frac{2}{M(M+1)}.
$$
\end{lemma}

See the Supplementary material for proofs of above lemmas.

The constant $B$ mostly depends on the data distribution in a feature space since the unnormalized vector $d$ indicates a degree of node corresponding to graph representing the dataset.
In the worst case, the bound for $Z(\bm{\alpha}^{(t)})$ scales as $B=\Theta(1/\sqrt{M})$ when all $M$ training samples collapse to a single point. 
This degenerate scenario is uninformative and rarely arises in practice. 
While we do not impose assumptions on the empirical data distribution, we do impose a prior on the parameter space from which $\bm{\alpha} \sim \mu$ is sampled. In particular, when $\bm{\alpha}$ is sampled from the Haar measure, we obtain a probabilistic upper bound on $B$. 

\begin{proposition}[Probabilistic Haar–random bound for $Z(\bm{\alpha}^{(t)})$]
Let $\bm{\alpha}^{(t)}\in\mathbb{C}^M$ be Haar–random on the unit sphere
$S^{2M-1}=\{\bm{u}\in\mathbb{C}^M:\|\bm{u}\|_2=1\}$, i.e., $\bm{\alpha}^{(t)}\sim\mathrm{Unif}(S^{2M-1})$.
Fix training data vectors $\{\bm{x_i}\}_{i=0}^{M-1}\subset\mathbb{C}^{2^n}$ with $\|\bm{x_i}\|_2=1$. 
Then for every $\delta\in(0,1)$,
$$
\Pr \left(\, \big|Z(\bm{\alpha}^{(t)})\big| \le  \frac{\ln(4/\delta)}{M}\,\right)
 \ge  1-\delta.
$$
\end{proposition}

\begin{proof}
Recall a standard concentration inequality on the complex unit sphere \cite{HighDimensionalProbability}:
for any fixed unit vector $\bm{v}\in\mathbb{C}^M$ and any $\varepsilon>0$,
\begin{align}\label{eq:sphere-conc}
\Pr \left(\, |\langle \bm{\alpha}^{(t)}, \bm{v}\rangle| > \varepsilon \,\right)
 \le  2\exp \big(- M\varepsilon^2\big),
\qquad \bm{\alpha}^{(t)}\sim\mathrm{Unif}(S^{2M-1}).
\end{align}
We allocate the failure budget $\delta$ equally to the two inner products appearing in $Z$.
For a fixed unit vector $\bm{v}$, setting the right-hand side of Equation \ref{eq:sphere-conc}
equal to $\delta/2$ and solving for $\varepsilon$ gives $\varepsilon=\sqrt{\ln(4/\delta)/M}$, so that
\begin{align}\label{eq:eps-delta}
\Pr\!\left(\,|\langle \bm{\alpha}^{(t)}, \bm{v}\rangle|\le \sqrt{\tfrac{\ln(4/\delta)}{M}}\,\right)\ge 1-\tfrac{\delta}{2}.
\end{align}
Applying Equation \ref{eq:eps-delta} with $\bm{v}=\bm{d}/\|\bm{d}\|_2$ yields
\begin{align}\label{eq:alpha-d}
\Pr\!\left(\,|\langle \bm{\alpha}^{(t)}, \bm{d}\rangle|\le \|\bm{d}\|_2\sqrt{\tfrac{\ln(4/\delta)}{M}}\,\right)\ge 1-\tfrac{\delta}{2},
\end{align}
and applying it with $\bm{v}=\bm{\vec{1}}/\|\bm{\vec{1}}\|_2$, together with $\|\bm{\vec{1}}\|_2=\sqrt{M}$, yields
\begin{align}\label{eq:one-alpha}
\Pr\!\left(\,|\langle \bm{\vec{1}}, \bm{\alpha}^{(t)}\rangle|\le \sqrt{\ln(4/\delta)}\,\right)\ge 1-\tfrac{\delta}{2}.
\end{align}
By the union bound, the events in Equation \ref{eq:alpha-d} and \ref{eq:one-alpha} hold simultaneously
with probability at least $1-\delta$. On this event, using
$d_i=\sum_{j=0}^{M-1}|\langle \bm{x_i}\mid \bm{x_j}\rangle|^2\le M$, hence
$\|\bm{d}\|_2=\sqrt{\sum_{i=0}^{M-1}d_i^2}\le\sqrt{\sum_{i=0}^{M-1}M^2}=M^{3/2}$, we obtain
\begin{align}
\big|Z(\bm{\alpha}^{(t)})\big|
=\frac{1}{M^{2}}\,|\langle \bm{\alpha}^{(t)}, \bm{d}\rangle|\,|\langle \bm{\vec{1}}, \bm{\alpha}^{(t)}\rangle|
\le \frac{1}{M^{2}}\cdot M^{3/2}\sqrt{\tfrac{\ln(4/\delta)}{M}}\cdot\sqrt{\ln(4/\delta)}
=\frac{\ln(4/\delta)}{M}.
\end{align}
Therefore $\Pr\!\big(|Z(\bm{\alpha}^{(t)})|\le \tfrac{\ln(4/\delta)}{M}\big)\ge 1-\delta$.
\end{proof}

Intuitively, fix any nonzero vector $\bm{v} \in \mathbb{C}^M$ and draw $\bm{\alpha}$ uniformly at random from the complex unit sphere (i.e., according to the Haar-induced distribution). In a high dimension, concentration of measure implies that a random direction is almost orthogonal to any fixed direction with overwhelming probability; equivalently, the event that $\bm{\alpha}$ exhibits a substantial alignment with $v$ is exponentially unlikely in $M$. 

Using a probabilistic bound on $Z\left(\bm{\alpha}^{(t)}\right)$, we introduce a multiplicative sampling overhead $r$ to ensure that the estimator of Equation \ref{eqn:proj_end} appearing in the objective Equation \ref{eq:objective-main} is reliable.

\begin{theorem*}[Shot complexity of the penalty]
Let $\{Z_i\}_{i=0}^{r-1}$ be i.i.d.\ copies of
$Z=\tfrac{1}{M^2}|\langle \bm d,\bm\alpha\rangle|\,|\langle \bm{\vec{1}},\bm\alpha\rangle|$,
$\bm\alpha\sim\mathrm{Unif}(S^{2M-1})$, with mean $\mu:=\mathbb E[Z]$ and the moment
bounds of Lemma~2. Fix $\delta_2\in(0,1)$, set $\delta_1=\exp(-\ln^{c}(1/\delta_2))$
for some $c>1$, and suppose the precision parameter satisfies
$$
s \in \Omega \left(\sqrt{\frac{\ln(1/\delta_2)}{\delta_2}}
\exp \Big(-\tfrac12\ln^{c}(1/\delta_2)\Big)\right)
\qquad(\delta_2\to0^+).
$$
Then, to guarantee
$$
\Pr\left(\Big|\frac1r\sum_{i=0}^{r-1}Z_i-\mu\Big|\le\frac{s}{M}\right)\ge 1-\delta_2,
$$
it suffices to take
$$
r \;=\; O\!\big(\delta_2\,\exp(\ln^{c}(1/\delta_2))\big).
$$
\end{theorem*}

\begin{proof}
By Proposition 1, $\Pr \big(|Z|>\tau\big) \leq \delta_1$ for
$\tau=\ln(4/\delta_1)/M$. Define the truncated variables
$\hat Z_i:=Z_i\,\mathbf 1\{|Z_i|\le\tau\}$, where $\mathbf 1\{\cdot\}$
denotes the indicator function, so that $0\le\hat Z_i\le\tau$ almost
surely, and let
\begin{align}
E_1:=\bigcap_{i=0}^{r-1}\{|Z_i|\le\tau\}=\bigcap_{i=0}^{r-1}\{Z_i=\hat Z_i\}.
\end{align}
By independence and Bernoulli's inequality,
$\Pr(E_1)\ge(1-\delta_1)^r\ge 1-r\delta_1$.
Since $0\le\hat Z_i\le Z_i$ pointwise, Lemma 2 gives
\begin{align}
\mathbb E[\hat Z]\le\mu\le\frac1M,
\qquad
\hat\sigma^2:=\mathrm{Var}[\hat Z]\le\mathbb E[\hat Z^2]\le\mathbb E[Z^2]\le\frac{2}{M(M+1)}.
\end{align}

The almost-sure bound on the centered variables is
\begin{align}
b:=\max_i|\hat Z_i-\mathbb E[\hat Z]|\le\tau+\mathbb E[\hat Z]
\le\frac{\ln(4/\delta_1)+1}{M}.
\end{align}

Now we consider the inequality of the gap between the variable $Z_i$ and the mean $\mu$ by decomposing about $E_1$,
\begin{align}
\Pr \left( \left| \frac{1}{r} \sum Z_i-\mu \right| \ge \frac{s}{M} \right)
&= \Pr \left( \left\{ \left| \frac{1}{r} \sum Z_i-\mu \right| \ge \frac{s}{M} \right\} \cap E_1 \right) + \Pr \left( \left\{ \left| \frac{1}{r} \sum Z_i-\mu \right| \ge \frac{s}{M} \right\} \cap E_1^{c} \right)\\
&\leq \Pr \left( \left| \frac{1}{r} \sum \hat{Z}_i-\mu \right| \ge \frac{s}{M} \right) + \Pr \left( E_1^c \right).
\end{align}
The equality is from $\Pr(A) = \Pr(A \cap B) + \Pr(A \cap B^{c})$, and the inequality is from the identity $Z_i = \hat{Z}_i$ on $E_1$ and $\Pr(A \cap B) \leq \Pr(A)$.
Let $\beta:=\mu-\mathbb E[\hat Z]=\mathbb E[Z\,\mathbf 1\{|Z|>\tau\}]\ge 0$ be the
truncation bias. Then if $\left| \frac{1}{r} \sum \hat{Z}_i - \mu \right| \geq \frac{s}{M}$,
\begin{align}
    \left| \frac{1}{r} \sum_i \hat{Z}_i - \mathbb{E}[\hat{Z}]\right|
    &= \left| \left(\frac{1}{r} \sum \hat{Z}_i - \mu\right) + \left( \mu - \mathbb{E}[\hat{Z}])\right) \right| \\
    &= \left| \left(\frac{1}{r} \sum \hat{Z}_i - \mu \right) + \beta \right| \\
    &\geq \left| \frac{1}{r} \sum_i \hat{Z}_i - \mu \right| - |\beta| \\
    &\geq \frac{s}{M} - \beta.
\end{align}
The first inequality is from the triangle inequality $|x+y| \geq |x| - |y|$. This indicates that 
\begin{align}
    &\left\{ \left| \frac{1}{r} \sum_i \hat{Z}_i - \mu \right|  \geq \frac{s}{M} \right\} \subseteq \left\{ \left| \frac{1}{r} \sum_i \hat{Z}_i - \mathbb{E}[\hat{Z}] \right| \geq \frac{s}{M} - \beta \right\} \\
    &\Rightarrow \Pr \left( \left| \frac{1}{r} \sum_i \hat{Z}_i - \mu \right| \geq \frac{s}{M} \right) \leq \Pr \left( \left| \frac{1}{r} \sum_i \hat{Z}_i - \mathbb{E}[\hat{Z}] \right| \geq \frac{s}{M} - \beta \right).
\end{align}
By Cauchy-Schwarz,
\begin{align}
\beta\le\sqrt{\mathbb E[Z^2]}\,\sqrt{\Pr(|Z|>\tau)}
\le\sqrt{\tfrac{2}{M(M+1)}}\,\sqrt{\delta_1}
=O\!\Big(\tfrac{\exp(-\frac12\ln^{c}(1/\delta_2))}{M}\Big).
\end{align}
Notice that for small $\delta_2$, $\frac{\beta}{s/M}=O \big(\sqrt{\delta_2/\ln(1/\delta_2)}\big)\to0$ and hence $\frac{s}{M}-\beta \ge \frac{s}{2M}$, and it suffices to control
$\Pr\!\big(|\tfrac1r\sum\hat Z_i-\mathbb E[\hat Z]|\ge \frac{s}{2M}\big)$. Hence, 
\begin{align}\label{eqn:proof_obj}
    \Pr \left( \left| \frac{1}{r} \sum Z_i-\mu \right| \ge \frac{s}{M} \right)
    &\leq \Pr \left( \left| \frac{1}{r} \sum \hat Z_i-\mathbb E[\hat Z] \right| \ge \frac{s}{2M} \right) + \Pr(E_1^{c}).
\end{align}

This motivates applying Bernstein's inequality on the truncated variables $\hat{Z}_i$. Recall that the $\hat Z_i$ are i.i.d., bounded by $\tau$, with variance $\hat\sigma^2$ and the
centered bound $b$. Bernstein's inequality at deviation $s/(2M)$ gives
\begin{align}
\Pr\!\Big(\big|\tfrac1r{\textstyle\sum}\hat Z_i-\mathbb E[\hat Z]\big|\ge\tfrac{s}{2M}\Big)
\le 2\exp\!\left(-\frac{r\,(s/2M)^2}{2\hat\sigma^2+\frac23 b\,(s/2M)}\right)
=: 2\exp\!\big(-r/A\big),
\quad
A:=\frac{2\hat\sigma^2+\frac23 b\,(s/2M)}{(s/2M)^2}.
\end{align}

Substituting $\hat\sigma^2\le\frac{2}{M(M+1)}$ and $b\le\frac{\ln(4/\delta_1)+1}{M}$,
\begin{align}
A=\frac{8\hat\sigma^2 M^2}{s^2}+\frac{4bM}{3s}
\le \frac{16M}{(M+1)s^2}+\frac{4(\ln(4/\delta_1)+1)}{3s}
\le \frac{16}{s^2}+\frac{4\big(\ln^{c}(1/\delta_2)+\ln 4+1\big)}{3s}.
\end{align}

Now we upper bound the right-hand side of Equation \ref{eqn:proof_obj} by allocating the failure budget $\delta_2$.
\begin{align}
\text{(I)}\quad 2\exp(-r/A)\le\tfrac{\delta_2}{2}
\iff r\ge A\ln\tfrac{4}{\delta_2};
\qquad
\text{(II)}\quad \Pr(E_1^{c}) \le r\delta_1\le\tfrac{\delta_2}{2}
\iff r\le\tfrac{\delta_2}{2\delta_1}.
\end{align}
Then, 
\begin{align}
\Pr\!\Big(\big|\tfrac1r{\textstyle\sum}Z_i-\mu\big|\ge\tfrac sM\Big)
\le \Pr(E_1^{c}) +2\exp(-r/A)
\le \frac{\delta_2}{2}+\frac{\delta_2}{2}=\delta_2.
\end{align}
(I) and (II) are simultaneously satisfiable if and only if
$A\ln(4/\delta_2)\le \delta_2/(2\delta_1)$, i.e.
\begin{align}
A\le \frac{\delta_2}{2\delta_1\ln(4/\delta_2)}
=\frac{\delta_2\,\exp(\ln^{c}(1/\delta_2))}{2\ln(4/\delta_2)}.
\end{align}

Enforcing this on the two terms of $A$ separately ($x \leq z/2$ and $y \leq z/2$ $\rightarrow$ $x+y \leq z$), the variance term
yields the binding constraint
\begin{align}
\frac{16}{s^2}\le \frac{\delta_2\,\exp(\ln^{c}(1/\delta_2))}{2\ln(4/\delta_2)}
\Longrightarrow 
s\ge 8\sqrt{\frac{\ln(4/\delta_2)}{\delta_2}}\;\exp\!\Big(-\tfrac12\ln^{c}(1/\delta_2)\Big),
\end{align}
which dominates the linear-term constraint as $\delta_2\to0^+$ and matches the hypothesis on $s$. At the boundary $A=\frac{\delta_2\,\exp(\ln^{c}(1/\delta_2))}{2\ln(4/\delta_2)}$,
\begin{align}
r=A\ln\tfrac{4}{\delta_2}=\frac{\delta_2\,\exp(\ln^{c}(1/\delta_2))}{2}
=O\!\big(\delta_2\,\exp(\ln^{c}(1/\delta_2))\big).
\end{align}
\end{proof}

We assumed that $\bm{\alpha}$ is Haar-random, since trainable quantum circuits are not guaranteed to follow any specific design in practice. However, the multiplicative sampling overhead $r$ for $\bm{\alpha}$ drawn from a unitary 2-design can also be obtained by invoking Welch's bound for complex projective 2-designs \cite{ComplexProj2designs} together with Markov's inequality applied to the fourth moment of $Z(\bm{\alpha}^{(t)})$. This yields the probabilistic bound in Proposition~1 as follows:
\begin{align}
    \operatorname{Pr}\left(\left|Z\left(\bm{\alpha}^{(t)}\right)\right| \leq \frac{(2 / \delta)^{1 / 2}}{M}\right) \geq 1-\delta,
\end{align}
preserving the inverse-linear dependence on $M$. Hence, the required multiplicative factor $r$ is unaffected with respect to $M$.

\color{black}
\section{Robustness heuristic during optimization}\label{app:robustness-heuristic}
During optimization, we maintain a buffer of the last ten penalty values. If the current penalty exceeds the maximum value in the buffer, the optimizer rejects the current step and reverts the parameters to the previous values. Once the penalty becomes exponentially small (we set $\varepsilon_{\text {threshold }}=10^{-8}$ in our simulation), the acceptance rule switches to a hard threshold: if the current penalty exceeds $\varepsilon_{\text {threshold }}$, the current step is rejected. This mechanism improves robustness to statistical noise by preventing the optimizer from moving far from the minimizer of the objective.

\newpage
\setcounter{section}{0}
\setcounter{figure}{0}
\setcounter{table}{0}
\setcounter{equation}{0}
\setcounter{theorem}{0}
\renewcommand{\thesection}{S\arabic{section}}
\renewcommand{\thefigure}{S\arabic{figure}}
\renewcommand{\thetable}{S\arabic{table}}
\renewcommand{\theequation}{S\arabic{equation}}
\renewcommand{\thetheorem}{S\arabic{theorem}}
\renewcommand{\theHsection}{S\arabic{section}}
\renewcommand{\theHfigure}{S\arabic{figure}}
\renewcommand{\theHtable}{S\arabic{table}}
\renewcommand{\theHequation}{S\arabic{equation}}
\renewcommand{\theHtheorem}{S\arabic{theorem}}
\let\sectionname\relax  
\allowdisplaybreaks

\begin{center}
  {\large\bfseries Supplementary Material for\\[0.3em]
  Quantum Spectral Clustering Framework via Compact Circuit Structures}\\[1.5em]
  Hyeong-gyu Kim\textsuperscript{1}, Siheon Park\textsuperscript{1},
  and June-Koo Kevin Rhee\textsuperscript{1,2,*}\\[0.5em]
  {\small\textsuperscript{1}School of Electrical Engineering, KAIST, Daejeon, Republic of Korea}\\
  {\small\textsuperscript{2}Qunova Computing, Inc., Daejeon, Republic of Korea}
\end{center}
\bigskip

\section{Spectral Clustering}\label{subsec2-1}

Spectral clustering is a well-studied classical unsupervised machine learning method that clusters datasets by modeling them as a graph \cite{TutorialSC}. Given a classical dataset $\mathcal{X}=\left\{x_0, \ldots, x_{M-1}\right\} \subset \mathbb{R}^d$, we can define a non-negative similarity $A_{i j}$ between each pair of data points $x_i, x_j$. Interpreting each data point $x_i$ as a vertex in the vertex set $V$ and the edge set $E$ as the weight of an edge between two vertices, we can define a weighted and undirected graph $G=(V, E)$ corresponding to the dataset $\mathcal{X}$.

Based on the similarities between data points, we can define the weighted adjacency matrix $A=\left(A_{i j}\right)_{i, j=0, \ldots, M-1}$ for the graph $G$. The degree of each vertex is defined as $d_i=\sum_{j=0}^{M-1} A_{i j}$ and using this, we can construct a diagonal matrix $D=\operatorname{diag}\left(d_0, \ldots, d_{M-1}\right)$, which is often referred as a degree matrix. One of the most well-known approximate spectral clustering methods for a dataset with two clusters involves defining the unnormalized graph Laplacian matrix $\mathcal{L}=D-A$ and applying Rayleigh-Ritz method to approximate eigenvalues. After computing the eigenvalues $\lambda_0 \leq \lambda_1 \leq \cdots \leq \lambda_{M-1}$ and the corresponding eigenvectors $v_0, v_1, ..., v_{M-1}$ of $\mathcal{L}$, the second smallest eigenvalue $\lambda_1$ and its corresponding eigenvector $v_1$, known as the Fiedler vector, are used to determine the clusters by examining the signs of the elements of the vector.

Inspired by this approach, instead of computing the whole eigenvector of the graph Laplacian matrix $\mathcal{L}$, we can approximate it by reformulating the problem into the Rayleigh quotient that converges approximately to the fiedler vector of $\mathcal{L}$:
\begin{align}
& \min _{\alpha \in \mathbb{R}^M} \alpha^T\left(\mathcal{L}+\xi v_0 v_0^T \right) \alpha \label{eqn:quad_cost}, \\ 
& \text { s.t. } \|\alpha\|^2_2=1. \nonumber
\end{align}

While it is a valid option to use the unnormalized Laplacian matrix as the objective function, we employ the normalized Laplacian matrix in the main text. This introduces a normalization of the matrix elements with respect to the degrees of the connected vertices, making it less likely that the trainable unitary need to express a strong bias as a solution.

\section{Efficient Weighted Sum Computation inspired from Swap-Test Classifier}
\begin{figure}[t!]
    \centering
    \begin{minipage}[t]{0.75\textwidth}
        \centering
        \includegraphics[width=\textwidth]{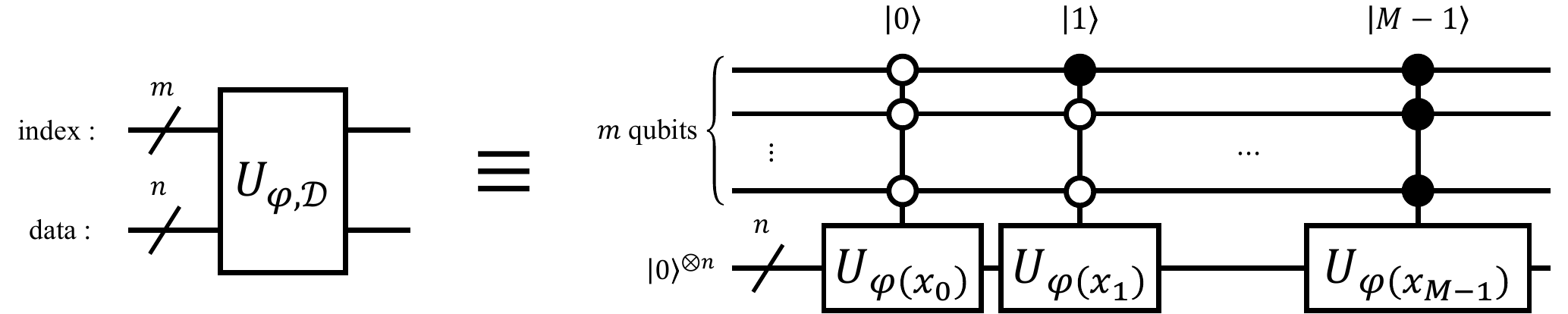}
        \label{fig:TrainingMultiplexor}
        \\
        \includegraphics[width=\textwidth]{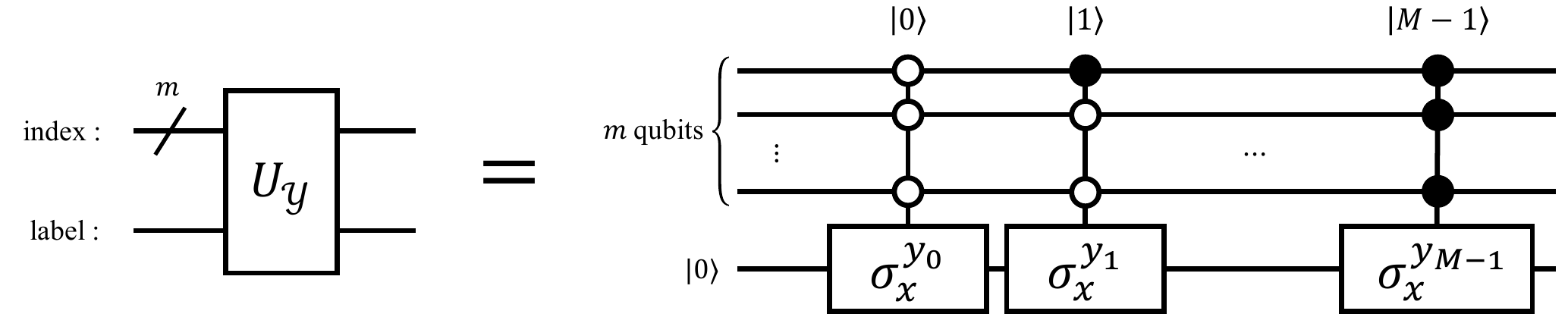}
        \subcaption{}
        \label{fig:TestMultiplexor}
    \end{minipage}
    \vspace{2ex}
    \begin{minipage}[t]{0.60\textwidth}
        \centering
        \includegraphics[width=\textwidth]{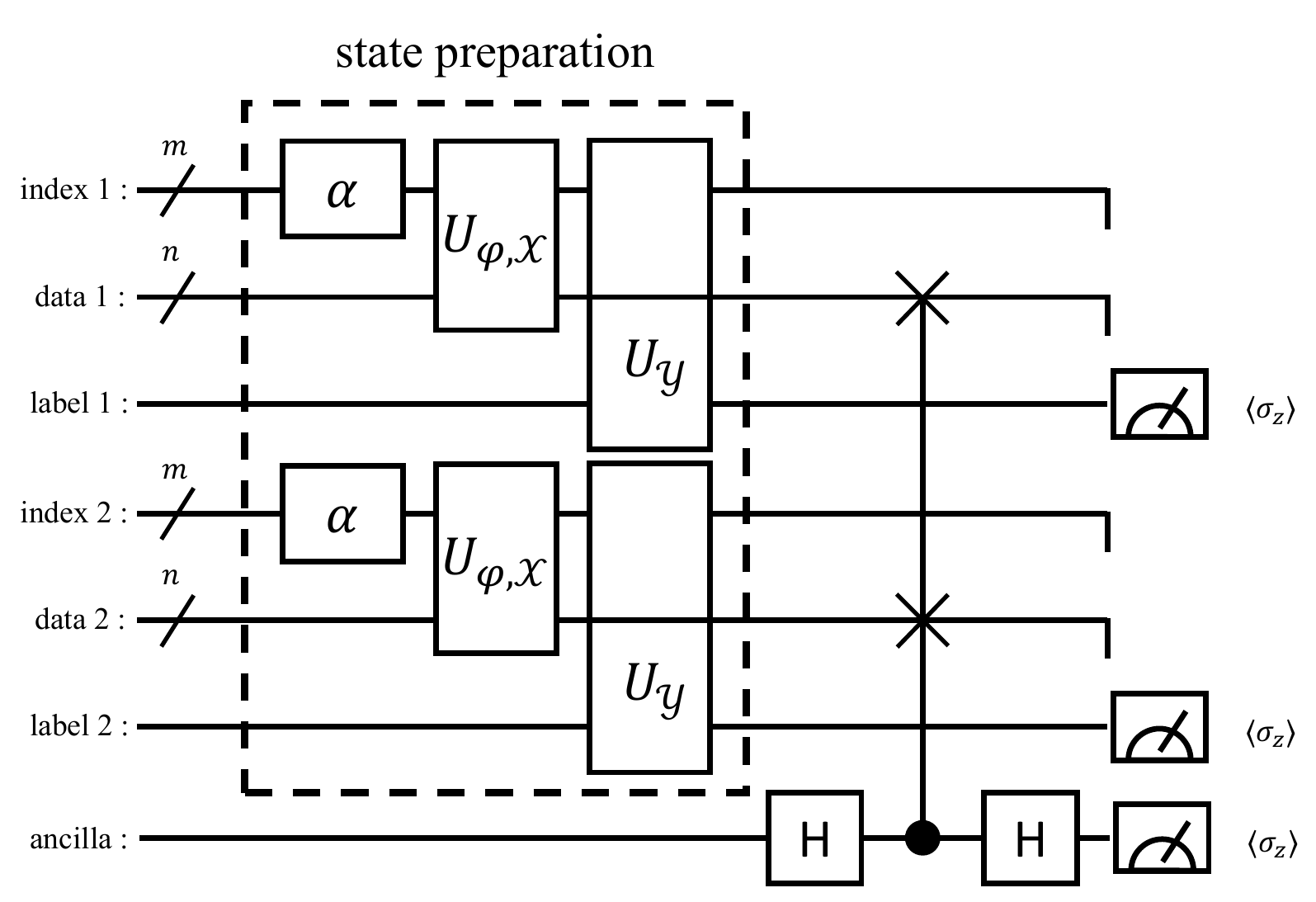}
        \subcaption{}
        \label{fig:EfficientWeight}
    \end{minipage}
    \caption{(a) Quantum circuit representation of the unitary blocks $U_{\varphi, \mathcal{X}}$ and $U_\mathcal{Y}$. (b) Quantum circuit diagram illustrating the efficient weighted sum computation. The measurement result in (b) corresponds to Eq.(\ref{eqn:efficient_weighted_sum_measurement}).}
    \label{fig:EfficientWeightedSumFigures}
\end{figure}
The Swap-Test Classifier (STC) is one of the family of quantum distance-based classifiers in quantum machine learning \cite{PatrickR, Mschuld, DKPark, CBlank}. Its key feature is that it can be implemented in the language of a quantum circuit with only a few operations, thereby circumventing the need for density matrix exponentiation \cite{QPCA} and the quantum linear system solving algorithm \cite{HHL}—issues that were present in the classifier model originally proposed in \cite{PatrickR}. Subsequently, with the introduction of a label qubit, it was further refined into the Hadamard classifier \cite{Mschuld} and the swap-test classifier \cite{CBlank, DKPark}, emphasizing efficient classifier implementation via minimal quantum circuits.

One interesting extension of the STC model is an efficient method for computing the weighted sum of kernel matrix elements. This approach was first introduced in \cite{SPark} to calculate the empirical risk of an approximated quantum support vector machine. By embedding a training dataset with uniformly controlled gates \cite{UniformlyControlledGate} and applying the swap-test, the weighted sum of the quantum kernel (Gram) matrix elements can be computed more efficiently than by the na\"ive classical construction of the dense kernel matrix, provided that the quantum feature map $\varphi: \mathbb{R}^d \rightarrow \mathbb{C}^N$ is realized by a hardware-efficient embedding scheme (see, for example, Figure~\ref{fig:simulations-embedding (U)} in the main text). In our work, this approach is further extended to include the weighted sum of matrix representatives of graph (e.g., the sum of graph degrees for a given weight vector), implemented via an efficient quantum circuit.

For the given training dataset $\left\{\left(x_i, y_i\right)\right\}_{i=0}^{M-1} \subset \mathcal{X} \times \mathcal{Y}$ and assuming a well-defined quantum state preparation process, STC assumes that a quantum state can be prepared as follows:
\begin{align}
|\Psi\rangle=\frac{1}{M} \sum_{i, j=0}^{M-1} \alpha_i \alpha_{j} |i\rangle |\varphi\left(x_i\right)\rangle  |y_i\rangle |j\rangle |\varphi\left(x_{j}\right)\rangle |y_{j}\rangle |0\rangle
\end{align}
Here, \(\alpha_i, \alpha_{j} \in \mathbb{C}^M \) denotes an arbitrary amplitude that can be prepared during the quantum state preparation process, and serves to generate a non-uniformly weighted kernel. The \(i, j\)-th training data \( x_i, x_{j} \) are embedded into the quantum state via quantum feature map $\varphi(\cdot)$, which are represented by the indices $|i\rangle, |j\rangle$ and training data $|\varphi(x_i)\rangle, |\varphi(x_{j})\rangle$, respectively. The qubits $|y_i\rangle, |y_{j}\rangle$, initialized with \(y_i, y_{j} \in \{0, 1\}\), represents the labels of the \(i, j\)-th training data. The last qubit indicates an ancilla qubit. Following the procedure described in \cite{CBlank}, a swap-test is performed on the training data register and test data register conditioned on an ancilla qubit; subsequently, the \(Z\)-expectation value is calculated from the measurement of the label qubit registers and ancilla qubit, yielding the following expression:
\begin{align}
    \operatorname{Tr}\left(\sigma_z^{\text{(label 1)}} \sigma_z^{\text{(label 2)}} \sigma_z^{\text{(ancilla)}} \left|\Psi\right\rangle\left\langle\Psi\right|\right)=\frac{1}{M^2} \sum_{i, j=0}^{M-1} \alpha_i \alpha_{j} y_i y_{j}\left|\left\langle\varphi\left(x_i\right) \mid \varphi\left(x_{j}\right)\right\rangle\right|^2 \label{eqn:efficient_weighted_sum_measurement}
\end{align}
From the perspective of quantum circuit implementation, the advantage of this approach lies in that the overall circuit depth is dependent on the chosen quantum feature map scheme in state preparation circuit block. Furthermore, the computation of the weighted sum for a given quantum state can be efficiently realized by applying controlled-swap gates a number of times that scales with the number of qubits to encoded data features. In main text, we primarily focused on the efficient implementation of the weighted sum calculation for the Laplacian matrix via the STC circuit. Assuming a data encoding scheme where the circuit depth increases linearly with $ M $ and $n$ qubits are used  to encoded data $x_i \in \mathbb{R}^d$, the overall STC circuit, which applies a uniformly controlled rotation gate \cite{UniformlyControlledGate} on the index qubit register and training data register, has a $ O(M n) $ depth.

\section{Weighted Kernel Principal Component Analysis for
Spectral Clustering}\label{subsec2-3}

Traditional principal components analysis (PCA) \cite{PCAPearson1901, PCAHotelling1933} can be extended to operate on feature-mapped points in a high-dimensional space—a method commonly referred to as kernel PCA. By incorporating additional weights into the formulation, kernel PCA can be further extended to Weighted kernel PCA (WPCA). In this section, we introduce the concept of WPCA, as originally studied and applied to spectral clustering in \cite{Suyken_SVMPCA, Suykens_WPCA}. We clarify that much of the material presented in the following introduction is borrowed from these prior studies.

First, we assume that all data are centered in the feature space. For a dataset $\mathcal{X}=\left\{x_i\right\}_{i=0}^{M-1}$ with $d$ features sampled from the training set, the data point $x$ is mapped to a high-dimensional feature space through a feature map function $\varphi(\cdot)$, and the projection of the feature space vector $\varphi(x)$ onto a target space via a vector $w \in \mathbb{R}^N$ is given by $w^T \varphi(x)$. We can now define a PCA problem in the target space as follows:
\begin{align}\label{eqn:maxvar}
\max _w \operatorname{Var}\left(w^T \varphi(x)\right)=\operatorname{Cov}\left(w^T \varphi(x), w^T \varphi(x)\right) \cong w^T C w.
\end{align}
where $C=\frac{1}{M} \sum_{i=0}^{M-1} \varphi\left(x_i\right) \varphi\left(x_i\right)^T$ \cite{Suyken_SVMPCA}. Fundamentally, this equation implies that the elements of $w$ cannot be centered around zero. From the structural risk minimization principle, Eq.(\ref{eqn:maxvar}) can be rewritten as a constrained optimization problem with a regularization term for $w$ :
\begin{align}\label{eqn:WPCA_cost (primal)}
& \max _{w, e} J_p(w, e)=\gamma \frac{1}{2} e^T e-\frac{1}{2} w^T w, \\
& \text { s.t. } e=\Phi w. \nonumber
\end{align}
where $\Phi=\left[\varphi\left(x_0\right)^T ; \varphi\left(x_1\right)^T ; \ldots ; \varphi\left(x_{M-1}\right)^T\right]$\cite{Suyken_SVMPCA}. 

This is a PCA problem with kernels that seeks to find a good $w$ that maximizes the variance in a specific target space, but can also be modified to include a matrix $B$ that assigns certain weights to each component of $e$, yielding the formulation known as `weighted kernel PCA' \cite{Suykens_WPCA}:
\begin{align}
& \max _{w, e} J_p(w, e)=\gamma \frac{1}{2} e^T B e-\frac{1}{2} w^T w, \label{eqn:WPCA_cost (primal2)} \\
& \text { s.t. } e=\Phi w. \nonumber
\end{align}

Using the KKT conditions, this problem can be expressed as a dual problem with a Lagrangian multiplier vector $\alpha$:
\begin{align}
B A \alpha=\lambda \alpha.
\end{align}

Here, $\gamma=1/\lambda$ is the regularization parameter, and $A=\Phi^T \Phi$ is the Gram matrix.

The dual formulation is an eigenvalue problem involving $B$, and we have the freedom to choose $B$. If we choose $B=D^{-1}$, we can reinterpret the problem as a random walk problem, which is one possible interpretation of spectral clustering \cite{Suykens_WPCA};
\begin{align}\label{eqn:WPCA_primal}
D^{-1} A \alpha=\lambda \alpha,
\end{align}

which corresponds to eigensovling problem of the random walk Laplacian matrix $L_{\text{rw}}=I - D^{-1}A$ and known to be equivalent to the symmetric Laplacian matrix $L_{\text{sym}} = I - \sqrt{D^{-1}} A \sqrt{D^{-1}}$.

\section{Reduction to the Objective Function in the Main Text}
The equation proposed in equation (\ref{eqn:WPCA_primal}), without a doubt, an eigensolving problem for a non-symmetric matrix. To make this problem into an approximate optimization problem that can be utilized on quantum devices, we first redefine the dual formulation to involve a Hermitian matrix:
\begin{align}
& D^{-1} A \alpha=\lambda \alpha, \\
& \sqrt{D^{-1}} A \alpha=\lambda \sqrt{D} \alpha, \\
& \sqrt{D^{-1}} A \sqrt{D^{-1}} q=\lambda q. \quad(q=\sqrt{D} \alpha)
\end{align}
Thus, we can now use the Rayleigh quotient to transform this into an optimization problem involving the Hermitian matrix $\sqrt{D^{-1}} A \sqrt{D^{-1}}$:
\begin{align}
\arg \max_q \frac{q^T \sqrt{D^{-1}} A \sqrt{D^{-1}} q}{q^T q}=v_{M-1}, \\
\max_q \frac{q^T \sqrt{D^{-1}} A \sqrt{D^{-1}} q}{q^T q}=\lambda_{M-1}.
\end{align}
One important property here is that, similar to the fact that the eigenvector $v_0$ corresponding to the smallest eigenvalue $\lambda_0$ of the original unnormalized Laplacian matrix is the constant one vector $\vec{1}$, the eigenvector $v_{M-1}$ corresponding to $\lambda_{M-1}$ in equation (\ref{eqn:WPCA_primal}) is also the constant one vector $\vec{1}$. Using this property, the problem can be transformed to find $v_{M-2}$, which contains all the information necessary for binary spectral clustering:
\begin{align}
\arg \max_{q, q \perp q_0} \frac{q^T \sqrt{D^{-1}} A \sqrt{D^{-1}} q}{q^T q}=v_{M-2}, \\
\max_{q, q \perp q_0} \frac{q^T \sqrt{D^{-1}} A \sqrt{D^{-1}} q}{q^T q}=\lambda_{M-2}.
\end{align}

where $q_0 = \sqrt{D}\vec{1}$. Additionally, due to the weight matrix $B=D^{-1}$ in equation (\ref{eqn:WPCA_cost (primal2)}), the eigenvalues always satisfy $0 \leq \lambda_0, \ldots \lambda_{M-1} \leq 1$. Therefore, inspired from equation (\ref{eqn:quad_cost}) provided in Supplementary Section~\ref{subsec2-1}, we can further transform the problem as follows:
\begin{align}
\arg \max_q \frac{q^T\left(\sqrt{D^{-1}} A \sqrt{D^{-1}}-\xi q_0 q_0^T\right) q}{q^T q} & =v_{M-2}, \\
\max_q \frac{q^T\left(\sqrt{D^{-1}} A \sqrt{D^{-1}}-\xi q_0 q_0^T\right) q}{q^T q} & =\lambda_{M-2}.
\end{align}
Here, $\xi$ is a positive constant. By substituting back $q=\sqrt{D} \alpha$ :
\begin{align}
\arg \max_\alpha \frac{\alpha^T A \alpha-\xi \alpha^T D \vec{1} \vec{1}^T D \alpha}{\alpha^T D \alpha}=v_{M-2}, \\
\max_\alpha \frac{\alpha^T A \alpha-\xi \alpha^T D \vec{1} \vec{1}^T D \alpha}{\alpha^T D \alpha}=\lambda_{M-2}.
\end{align}

Now, without loss of generality, we extend $\alpha \in \mathbb{C}^M$ and redefine the Gram matrix $A$ based on the square of the absolute value of the inner product between feature space vectors, resulting in the final optimization problem:
\begin{align}
\arg \max_\alpha \frac{\alpha^{\dagger} A \alpha-\xi \alpha^{\dagger} D \vec{1} \vec{1}^T D \alpha}{\alpha^{\dagger} D \alpha}=v_{M-2}, \\ 
\max_\alpha \frac{\alpha^{\dagger} A \alpha-\xi \alpha^{\dagger} D \vec{1} \vec{1}^T D \alpha}{\alpha^{\dagger} D \alpha}=\lambda_{M-2}.
\end{align}
\newpage

\section{Proofs of Circuits in Main Text}\label{appendix1}
We use the complex number notation $\alpha_i$ to represent the $i$-th element of $U_{\alpha} |0\rangle^{\otimes m} = \sum_{i=0}^{M-1} \alpha_i |i\rangle$ in standard basis, which can also be expressed as $\alpha_i=\langle i | U_{\alpha} | 0^m \rangle$. We denote the quantum encoded state $\left|\varphi\left(x_i\right)\right\rangle$ as $\left|x_i \right\rangle$ and omit superscript $m$ in $|0^m\rangle$ for brevity. 

\subsection{Proof of $q_a(\alpha)$}
\begin{align}
&|0\rangle_{A_1} |0\rangle_{B_1} |0\rangle_{A_2} |0\rangle_{B_2} \\
\xrightarrow{{U_{\alpha}}_{(A_1)}} &\sum_{i=0}^{M-1} \alpha_i |i\rangle_{A_1} |0\rangle_{B_1} |0\rangle_{A_2} |0\rangle_{B_2} \\
\xrightarrow{\text{CNOT}_{(A_2 | A_1)}} &\sum_{i=0}^{M-1} \alpha_i |i\rangle_{A_1} |0\rangle_{B_1} |i\rangle_{A_2} |0\rangle_{B_2} \\
\xrightarrow{\left(U_{\varphi, \mathcal{D}}\right)_{(A_1,B_1)}, \left(\overline{U}_{\varphi, \mathcal{D}}\right)_{(A_2,B_2)}} &\sum_{i=0}^{M-1} \alpha_i |i\rangle_{A_1} |x_i\rangle_{B_1} |i\rangle_{A_2} |\overline{x}_i\rangle_{B_2} \\ 
\xrightarrow{\text{CNOT}_{(A_2 | A_1)}} &\sum_{i=0}^{M-1} \alpha_i |i\rangle_{A_1} |x_i\rangle_{B_1} |0\rangle_{A_2} |\overline{x}_i\rangle_{B_2} \\ 
\xrightarrow{\text{Tr}_{(B_1, A_2, B_2)}} &\sum_{i,i'=0}^{M-1} \alpha_i \overline{\alpha}_{i'} \langle x_{i'}|x_i\rangle  \langle \overline{x}_{i'}|\overline{x}_i\rangle  |i\rangle \langle i'|_{A_1} \\ 
= &\sum_{i,i'=0}^{M-1} \alpha_i \overline{\alpha}_{i'} \langle  x_{i'}|x_i\rangle  \overline{\langle x_{i'}|x_i\rangle} |i\rangle \langle i'|_{A_1} \\ 
= &\sum_{i,i'=0}^{M-1} \alpha_i \overline{\alpha}_{i'} \left|\langle x_i|x_{i'}\rangle\right|^2 |i\rangle \langle i'|_{A_1} \\ 
\xrightarrow{H_{(A_1)},\ M_{{00...0}_{(A_1)}}} &\frac{1}{M} \cdot \sum_{i,i'=0}^{M-1} \alpha_i \overline{\alpha}_{i'} \left|\langle x_i|x_{i'}\rangle\right|^2 &&
\end{align}

\newpage
\subsection{Proof of $q_d(\alpha)$}
\begin{align}
&|0\rangle_{A_1} |0\rangle_{B_1} |0\rangle_{A_2} |0\rangle_{B_1} |0\rangle_C \\
\xrightarrow{U_{\alpha_{(A_1)}}, H_{(A_2)}} \frac{1}{\sqrt{M}}&\sum_{i,j=0}^{M-1}  \alpha_i |i\rangle_{A_1} |0\rangle_{B_1} |j\rangle_{A_2} |0\rangle_{B_1} |0\rangle_C \\
\xrightarrow{\left(U_{\varphi, \mathcal{D}}\right)_{(A_1,B_1)}, \left(U_{\varphi, \mathcal{D}}\right)_{(A_2,B_2)}} \frac{1}{\sqrt{M}}&\sum_{i,j=0}^{M-1} \alpha_i |i\rangle_{A_1} |x_i\rangle_{B_1} |j\rangle_{A_2} |x_j\rangle_{B_2} |0\rangle_{C} \\ 
\xrightarrow{H_C \cdot \text{CSWAP}_{(B_1,B_2|C)} \cdot H_C} \frac{1}{2\cdot \sqrt{M}}&\sum_{i,j=0}^{M-1} \alpha_i \left(|i\rangle_{A_1} |x_i\rangle_{B_1} |j\rangle_{A_2} |x_j\rangle_{B_2} + |i\rangle_{A_1} |x_j\rangle_{B_1} |j\rangle_{A_2} |x_i\rangle_{B_2} \right) |0\rangle_{C} \\
&+ \left(|i\rangle_{A_1} |x_i\rangle_{B_1} |j\rangle_{A_2} |x_j\rangle_{B_2} - |i\rangle_{A_1} |x_j\rangle_{B_1} |j\rangle_{A_2} |x_i\rangle_{B_2} \right) |1\rangle_{C} \nonumber \\ 
\xrightarrow{\text{Tr}_{\left(A_1, A_2\right)} \cdot \langle \sigma_z \rangle_C} \frac{1}{2M}&\sum_{i,j=0}^{M-1} \alpha_i \overline{\alpha}_i \cdot \left(|x_i\rangle\langle x_j|_{B_1} \otimes |x_j\rangle\langle x_i|_{B_2} + |x_j\rangle\langle x_i|_{B_1} \otimes |x_i\rangle\langle x_j|_{B_2} \right)\\
\xrightarrow{\text{Tr}_{\left(B_1, B_2\right)}} \frac{1}{M}&\sum_{i,j=0}^{M-1} \alpha_i \overline{\alpha}_i \left|\langle x_i|x_j\rangle\right|^2 \cdot \\
= \frac{1}{M}&\sum_{i,j=0}^{M-1} |\alpha_i|^2 \left|\langle x_i|x_j\rangle\right|^2 \cdot &&
\end{align}

\newpage
\subsection{Proof of $q_p(\alpha)$}\label{app:circuits-c}
\begin{align}
&|0\rangle_{A_0} |0\rangle_{A_1} |0\rangle_{B_1} |0\rangle_{A_2} |0\rangle_{B_2} |0\rangle_{C} \\
\xrightarrow{\left(U_{\alpha}\right)_{(A_0)}, \left(H\right)_{(A_1)}} &\frac{1}{\sqrt{M}} \sum_{i,j=0}^{M-1} \overline{\alpha}_i |i\rangle_{A_0} |j\rangle_{A_1} |0\rangle_{B_1} |0\rangle_{A_2} |0\rangle_{B_2} |0\rangle_{C} \\
\xrightarrow{\text{CNOT}_{(A_2 | A_1)}} &\frac{1}{\sqrt{M}} \sum_{i,j=0}^{M-1} \overline{\alpha}_i |i\rangle_{A_0} |j\rangle_{A_1} |0\rangle_{B_1} |j\rangle_{A_2} |0\rangle_{B_2} |0\rangle_{C} \\
\xrightarrow{\left(U_{\varphi, \mathcal{D}}\right)_{(A_1,B_1)}, \left(\overline{U}_{\varphi, \mathcal{D}}\right)_{(A_2,B_2)}} &\frac{1}{\sqrt{M}}\sum_{i,j=0}^{M-1} \overline{\alpha}_i |i\rangle_{A_0} |j\rangle_{A_1} |x_j\rangle_{B_1} |j\rangle_{A_2} |\overline{x}_j\rangle_{B_2} |0\rangle_{C} \\ 
\xrightarrow{\text{CNOT}_{(A_2 | A_1)}} &\frac{1}{\sqrt{M}} \sum_{i,j=0}^{M-1} \overline{\alpha}_i |i\rangle_{A_0} |j\rangle_{A_1} |x_j\rangle_{B_1} |0\rangle_{A_2} |\overline{x}_j\rangle_{B_2} |0\rangle_{C} \\
\xrightarrow{\operatorname{Tr}(B_1, A_2, B_2)} &\frac{1}{M}\sum_{i,i',j,j'=0}^{M-1} \overline{\alpha}_i \alpha_{i'} \cdot \left|\langle x_j \mid x_{j'} \rangle \right|^2 \left( |i\rangle\langle i'|_{A_0} \otimes |j\rangle \langle j'|_{A_1} \otimes |0\rangle\langle 0|_{C}\right)\\
\xrightarrow{H_C \cdot \text{CSWAP}_{(A_0, A_1|C)}} \frac{1}{2M}&\sum_{i,i',j,j'=0}^{M-1} \left( \overline{\alpha}_i \alpha_{i'} \cdot \left|\langle x_j \mid x_{j'} \rangle \right|^2 \right) \\
& \big( |i\rangle \langle i'|_{A_0} \otimes |j\rangle \langle j'|_{A_1} \otimes |0\rangle \langle 0|_C + |j\rangle \langle i'|_{A_0} \otimes |i\rangle \langle j'|_{A_1} \otimes |1 \rangle \langle 0|_C \nonumber \\
& + |i\rangle \langle j'|_{A_0} \otimes |j\rangle \langle i'|_{A_1} \otimes |0\rangle \langle 1|_C + |j\rangle \langle j'|_{A_0} \otimes |i \rangle \langle i'|_{A_1} \otimes |1\rangle \langle 1|_C \big) \nonumber \\
\xrightarrow{H_{A_0} \cdot \langle M_{00..0} \rangle_{A_0}} \frac{1}{2M^2}&\sum_{i,i',j,j'=0}^{M-1} \left( \overline{\alpha}_i \alpha_{i'}  \cdot \left|\langle x_j \mid x_{j'} \rangle \right|^2 \right) \\
& \big( |j\rangle \langle j'|_{A_1} \otimes |0\rangle \langle 0|_C + |i\rangle \langle j'|_{A_1} \otimes |1 \rangle \langle 0|_C \\
&+ |j\rangle \langle i'|_{A_1} \otimes |0\rangle \langle 1|_C + |i \rangle \langle i'|_{A_1} \otimes |1\rangle \langle 1|_C \big) \nonumber \\
\xrightarrow{\operatorname{Tr}_{A_1} \cdot \langle \sigma_x \rangle_C} \frac{1}{2M^2} &\sum_{i,i',j,j'=0}^{M-1} \left( \overline{\alpha}_i \alpha_{i'} \cdot \left|\langle x_j \mid x_{i} \rangle \right|^2 \right) + \left( \overline{\alpha}_i \alpha_{j} \cdot \left|\langle x_j \mid x_{j'} \rangle \right|^2 \right) = \frac{1}{M^2} \operatorname{Re } \overline{\alpha}_i \alpha_j \cdot \left|\langle x_j \mid x_{j'} \rangle \right|^2\\
\xrightarrow{\operatorname{Tr}_{A_1} \cdot \langle \sigma_y \rangle_C} \frac{1}{2M^2} &\sum_{i,i',j,j'=0}^{M-1} \left( \overline{\alpha}_i \alpha_{i'} \cdot \left|\langle x_j \mid x_{i} \rangle \right|^2 \right) - \left( \overline{\alpha}_i \alpha_{j} \cdot \left|\langle x_j \mid x_{j'} \rangle \right|^2 \right) = \frac{1}{M^2} \operatorname{Im } \overline{\alpha}_i \alpha_j \cdot \left|\langle x_j \mid x_{j'} \rangle \right|^2 \nonumber
\end{align}
\newpage
\subsection{Proof of Eqs.~(\ref{eqn:score_end1})--(\ref{eqn:score_end2})}\label{app:circuits-d}
\begin{align}
\frac{1}{\sqrt{M}} &|0\rangle_{C_1} |0\rangle_{A_1} |0\rangle_{B_1} |0\rangle_{B_2} |0\rangle_{C_2} \\
\xrightarrow{(Hadamard-Test)_{(A_1|C_1)}} \frac{1}{\sqrt{2}}&\sum_{i=0}^{M-1} \frac{1}{\sqrt{M}} |0\rangle_{C_1} |i\rangle_{A_1} |0\rangle_{B_1} |0\rangle_{B_2} |0\rangle_{C_2} + \alpha_i |1\rangle_{C_1} |i\rangle_{A_1} |0\rangle_{B_1} |0\rangle_{B_2} |0\rangle_{C_2} \\
\xrightarrow{(U_{\varphi, \mathcal{D}})_{(A_1, B_2)}, U_{(\varphi, \hat{x}})_{B_2}} \frac{1}{\sqrt{2}}& \sum_{i=0}^{M-1} \frac{1}{\sqrt{M}} |0\rangle_{C_1} |i\rangle_{A_1} |x_i\rangle_{B_1} |\hat{x}\rangle_{B_2} |0\rangle_{C_2} + \alpha_i |1\rangle_{C_1} |i\rangle_{A_1} |x_i\rangle_{B_1} |\hat{x} \rangle_{B_2} |0\rangle_{C_2}\\
\xrightarrow{H_{C_2} \cdot \text{CSWAP}_{(B_1,B_2|C_2)} \cdot H_{C_2}} \frac{1}{2}&\sum_{i=0}^{M-1} \left(\frac{1}{\sqrt{M}}|0\rangle_{C_1} |i\rangle_{A_1} |x_i\rangle_{B_1} |\hat{x}\rangle_{B_2} + \alpha_i|1\rangle_{C_1} |i\rangle_{A_1} |x_i\rangle_{B_1} |\hat{x}\rangle_{B_2} \right.\\ 
&\left.+\frac{1}{\sqrt{M}}|0\rangle_{C_1} |i\rangle_{A_1} |\hat{x}\rangle_{B_1} |x_i\rangle_{B_2} + \alpha_i|1\rangle_{C_1} |i\rangle_{A_1} |\hat{x}\rangle_{B_1} |x_i\rangle_{B_2} \right)|0\rangle_{C_2} \\
&+ \left(\frac{1}{\sqrt{M}}|0\rangle_{C_1} |i\rangle_{A_1} |x_i\rangle_{B_1} |\hat{x}\rangle_{B_2} + \alpha_i |1\rangle_{C_1} |i\rangle_{A_1} |x_i\rangle_{B_1} |\hat{x}\rangle_{B_2} \right. \\
&\left.-\frac{1}{\sqrt{M}}|0\rangle_{C_1} |i\rangle_{A_1} |\hat{x}\rangle_{B_1} |x_i\rangle_{B_2} - \alpha_i|1\rangle_{C_1} |i\rangle_{A_1} |\hat{x}\rangle_{B_1} |x_i\rangle_{B_2} \right)|1\rangle_{C_2} \\
\xrightarrow{\text{Tr}_{(A_1, B_1, B_2)} \cdot \langle \sigma_z \rangle_{C_2}} \frac{1}{2} &\sum_{i=0}^{M-1} \left(\frac{1}{M}|0\rangle \langle 0|_{C_1} + \frac{\overline{\alpha_i}}{\sqrt{M}}|0\rangle \langle 1|_{C_1} + \frac{\alpha_i}{\sqrt{M}}|1\rangle \langle 0|_{C_1} + \alpha_i \overline{\alpha}_i |1\rangle \langle 1|_{C_1}\right) \left|\langle x_i|\hat{x}\rangle\right|^2 \\
&+ \left(\frac{1}{M}|0\rangle \langle 0|_{C_1} - \frac{\overline{\alpha_i}}{\sqrt{M}}|0\rangle \langle 1|_{C_1} - \frac{\alpha_i}{\sqrt{M}}|1\rangle \langle 0|_{C_1} + \alpha_i \overline{\alpha}_i |1\rangle \langle 1|_{C_1}\right) \left|\langle x_i|\hat{x}\rangle\right|^2 \nonumber \\
\xrightarrow{\langle \sigma_x \rangle_{C_1}} \frac{1}{2\sqrt{M}} &\sum_{i=0}^{M-1} |\langle x_i | \hat{x}\rangle |^2 (\alpha_i + \overline{\alpha}_i) = \frac{1}{\sqrt{M}} \sum_{i=0}^{M-1}  |\langle x_i | \hat{x}\rangle |^2 \cdot \sqrt{|\alpha_i|^2} \cos(\phi_i)\\
\xrightarrow{\langle \sigma_y \rangle_{C_1}} \frac{1}{2\sqrt{M}} &\sum_{i=0}^{M-1} |\langle x_i | \hat{x}\rangle |^2 (\alpha_i - \overline{\alpha}_i) = -\frac{1}{\sqrt{M}} \sum_{i=0}^{M-1}  |\langle x_i | \hat{x}\rangle |^2 \cdot \sqrt{|\alpha_i|^2} \sin(\phi_i) \nonumber
\end{align}

The sign difference between $q_{H_x}$ and $q_{H_y}$ can be ignored since the phase in our scheme is relative.

\newpage
\section{Convergence of the main objective function}
\begin{proposition}[Complex extension of proposition 4.6.1 from \cite{OptimizationOnMatrixManifolds}]
    Let $A=A^\dagger$ be an $M \times M$ Hermitian matrix. A unit-norm vector $\alpha \in \mathbb{C}^M$ is an eigenvector of $A$ if and only if it is a critical point of the Rayleigh quotient.
\end{proposition}

\begin{proposition}[Complex extension of proposition 4.6.2 from \cite{OptimizationOnMatrixManifolds}]
    Let $A=A^\dagger$ be an $M \times M$ Hermitian matrix with eigenvalues $\lambda_1 \leq \cdots \leq \lambda_M$ and associated orthonormal eigenvectors $v_1, \ldots, v_M$. Then
    \begin{enumerate}
        \item $\pm v_1$ are local and global minimizers of the Rayleigh quotient; if the eigenvalue $\lambda_1$ is simple, then they are the only minimizers.
        \item $\pm v_M$ are local and global maximizers of the Rayleigh quotient; if the eigenvalue $\lambda_M$ is simple, then they are the only maximizers.
        \item $\pm v_q$ corresponding to interior eigenvalues (i.e., strictly larger than $\lambda_1$ and strictly smaller than $\lambda_M$ ) are saddle points of the Rayleigh quotient.
    \end{enumerate}
\end{proposition}

This provides a guarantee that a stable convergence of descent method to the smallest eigenvector of the Rayleigh quotient form of the Laplacian matrix while the other critical points remains the optimizer unstable. From the orthogonality of eigenvectors, $\alpha$ becomes orthogonal to $\bm{d}$ ($=\bm{D\vec{1}}$) when $Z(\alpha)$ converges to its global minimum. Therefore, we can trivially assume that the $Z(\alpha)$ converges to 0 if the objective function given in the main text converges to its global maximum with high probability. Based on this, we introduced the iteration $t$ as a indicator for the number of iteration during the optimization step that converges to the global minimum when $t \rightarrow \infty$.

\section{Proof of Lemma 2}
\begin{itemize}
    \item (Proof of convergence of $\mathbb{E}[Z(\alpha^{(t)})]$)
    (For brevity, we write $Z(\alpha^{(t)})$ as $Z^t$) Fix any $\varepsilon>0$. Using the indicator function $\mathbf{1}_{\{\cdot\}}$,
$$
Z^t
= Z^t \mathbf{1}_{\{Z^t\le \varepsilon\}}
  + Z^t \mathbf{1}_{\{Z^t> \varepsilon\}}.
$$
By linearity of expectation and the bound $0\le Z^t\le B$,
$$
\mathbb{E}[Z^t]
= \mathbb{E}\!\left[ Z^t \mathbf{1}_{\{Z^t\le \varepsilon\}}\right]
 + \mathbb{E}\!\left[ Z^t \mathbf{1}_{\{Z^t> \varepsilon\}}\right]
\le \varepsilon + B \cdot \Pr(Z^t>\varepsilon).
$$
Since $Z^t \xrightarrow{\mathbb{P}} 0$, we have $\Pr(Z^t>\varepsilon)\to 0$, hence
$$
\limsup_{t\to \infty}\mathbb{E}[Z^t]  \le  \varepsilon.
$$
Because $\varepsilon>0$ was arbitrary, $\limsup_{t\to \infty}\mathbb{E}[Z^t]\le 0$. On the other hand, $Z^t\ge 0$ implies $\liminf_{t\to \infty}\mathbb{E}[Z^t]\ge 0$. Therefore,
$$
\lim_{t\to \infty}\mathbb{E}[Z^t]=0.
$$
    \item (Proof of convergence of $\mathrm{Var}[Z(\alpha^{(t)})]$) 
By the continuous mapping theorem, $(Z^t)^2 \xrightarrow{\mathbb{P}} 0$ as $t\rightarrow \infty$. 
Since $0\le (Z^t)^2 \le B^2$, the same argument (applied to $(Z^t)^2$) gives
$\mathbb{E}[(Z^t)^2]\to 0$. Therefore, for $t \rightarrow \infty$,
$$
\mathrm{Var}(Z^t)
=\mathbb{E}\!\left[(Z^t)^2\right] - \big(\mathbb{E}[Z^t]\big)^2
 \longrightarrow  0-0  =  0.
$$
\end{itemize}
 
\section{Proof of Lemma 3}
\begin{itemize}
    \item (Proof of the bound of $\mathbb{E}[Z]$) By Cauchy--Schwarz on the underlying probability space,
$$
\mathbb{E}\!\left[|\langle d,\alpha\rangle|\,|\langle \vec{1},\alpha\rangle|\right]
 \le \sqrt{\mathbb{E}|\langle d,\alpha\rangle|^2 \mathbb{E}|\langle \vec{1},\alpha\rangle|^2}.
$$
Haar isotropy gives $\mathbb{E}|\langle v,\alpha\rangle|^2=\|v\|_2^2/M$ for all $v\in\mathbb{C}^M$.
Therefore
$$
\mathbb{E}\!\left[|\langle d,\alpha\rangle|\,|\langle \vec{1},\alpha\rangle|\right]
 \le \sqrt{\frac{\|d\|_2^2}{M}\cdot\frac{\|\vec{1}\|_2^2}{M}}
 = \frac{\|d\|_2\,\|\vec{1}\|_2}{M}.
$$
Multiplying by $1/M^2$ and using $\|d\|_2 \leq M^{3/2}$, $\|1\|_2=\sqrt{M}$ yields $\mathbb{E}[Z] \leq 1/M$.
    \item (Proof of the bound of $\mathrm{Var}[Z]$) We have $\mathrm{Var}(Z)=\mathbb{E}[Z^2]-(\mathbb{E}Z)^2\le \mathbb{E}[Z^2]$.

Compute $\mathbb{E}[Z^2]$ using the complex projective $2$-design identity
$$
\mathbb{E}\!\left[(|\alpha\rangle\langle\alpha|)^{\otimes 2}\right]
=\frac{I+F}{M(M+1)},
$$
where $F$ swaps the two tensor factors. Writing
$|\langle d,\alpha\rangle|^2=\alpha^*(dd^*)\alpha$ and
$|\langle \vec{1},\alpha\rangle|^2=\alpha^*(\vec{1}\vec{1}^*)\alpha$, one obtains the standard quadratic form moment formula
$$
\mathbb{E}\!\left[(\alpha^*A\alpha)(\alpha^*B\alpha)\right]
=\frac{\operatorname{tr}A\,\operatorname{tr}B+\operatorname{tr}(AB)}{M(M+1)}
\qquad(\text{for Hermitian }A,B).
$$
Applying this with $A=dd^*$ and $B=\vec{1}\vec{1}^*$ yields
$$
\mathbb{E}\!\left[|\langle d,\alpha\rangle|^2\,|\langle \vec{1},\alpha\rangle|^2\right]
=\frac{\|d\|_2^2\,\|\vec{1}\|_2^2+|\langle d,\vec{1}\rangle|^2}{M(M+1)}
$$
Therefore
$$
\mathbb{E}[Z^2]
=\frac{1}{M^4}\,
\mathbb{E}\!\left[|\langle d,\alpha\rangle|^2\,|\langle \vec{1},\alpha\rangle|^2\right]
=\frac{M\|d\|_2^2+|\langle d,\vec{1}\rangle|^2}{M^5(M+1)}
\le \frac{2}{M(M+1)},
$$
\end{itemize}

\newpage
\section{Additional Results}
\subsection{Empirical Comparison to the Classical Baselines}
\begin{figure}[t]
    \centering
    \begin{minipage}[t]{0.49\textwidth}
        \centering
        \includegraphics[width=\textwidth]{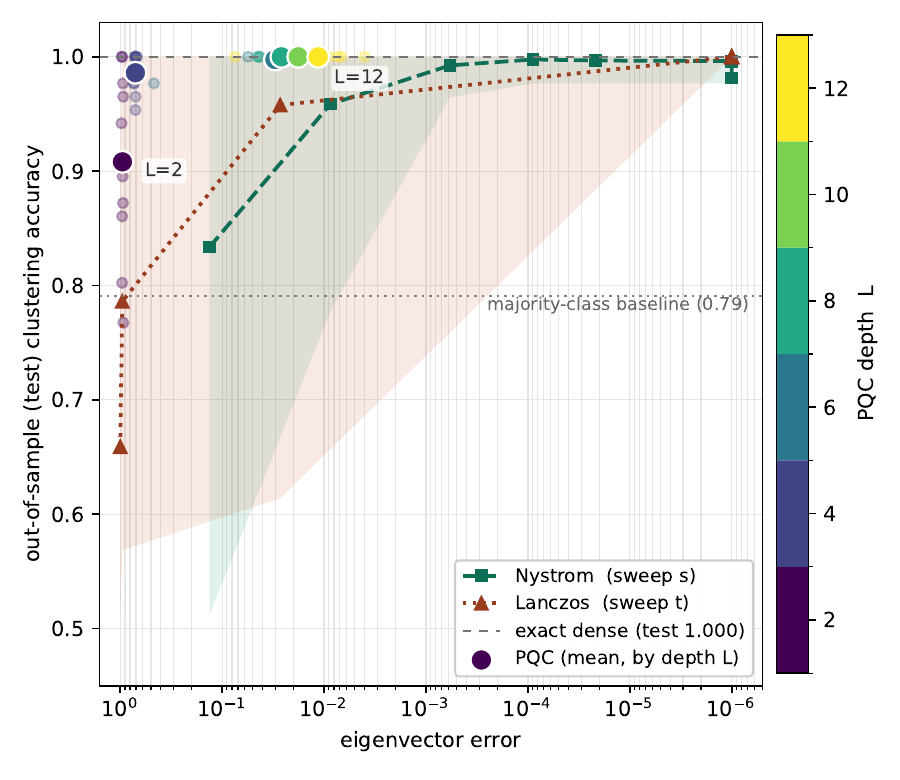}
        \subcaption{\textit{Iris} ($M=64$)}
        \label{fig:supp-oos-iris}
    \end{minipage}\hfill
    \begin{minipage}[t]{0.49\textwidth}
        \centering
        \includegraphics[width=\textwidth]{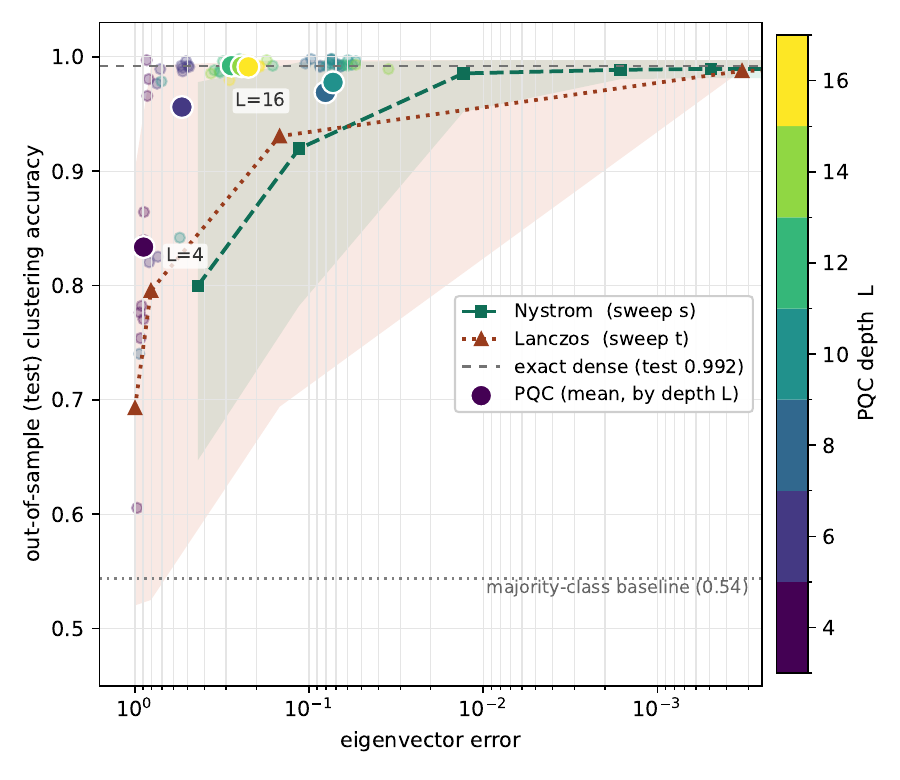}
        \subcaption{\textit{MNIST} ($M=128$)}
        \label{fig:supp-oos-mnist}
    \end{minipage}
    \caption{Held-out clustering accuracy versus eigenvector-approximation quality on a shared axis. The $x$-axis is the error
    $1-|\langle \hat{\bm v},\bm v^\ast\rangle|^2$ to the exact second eigenvector
    $\bm v^\ast$ of $\bm M=\bm D^{-1/2}\bm A\bm D^{-1/2}$ (finer to the right); the
    $y$-axis is test accuracy under the same score
    $s(\hat{\bm x})=\sum_i \hat v_i\,\kappa(\bm x_i,\hat{\bm x})$ for all methods,
    so only the recovered eigenvector differs. The parameterized quantum circuit is swept over depth $L$,
    Nystr\"om over its landmark count $s$, and Lanczos over its
    iteration count $t$; curves are the mean over ten splits with a $5$--$95$
    percentile band, the dashed line the exact-dense ceiling, the dotted line the
    majority-class baseline.}
    \label{fig:supp-oos}
\end{figure}
To complement Section~\ref{sec:comparison}, we compare the proposed estimator against two scalable classical methods, Nystr\"om and Lanczos, out-of-sample on the held-out test splits. All three target the same object - the second eigenvector $\bm v^\ast$ of the symmetric-normalized affinity $\bm M=\bm D^{-1/2}\bm A\bm D^{-1/2}$ - and are scored with the identical rule $s(\hat{\bm x})=\sum_i \hat v_i\,\kappa(\bm x_i,\hat{\bm x})$ used elsewhere. With this common scorer, the only quantity that differs is the quality of the recovered eigenvector, plotted on the horizontal axis of Figure~\ref{fig:supp-oos}.
 
For Nystr\"om, both the training eigenvector and its test extension are built from the same $s$ uniformly sampled landmark columns; Lanczos returns the Ritz vector of the second-largest Ritz value after $t$ iterations with full reorthogonalization. We sweep $s\in\{2,4,8,12,16,24,32,48,64\}$ and $t\in\{2,3,4,6,8,10,12,16,24,32,64\}$, up to the full set size $M$, where both recover the exact eigenvector. The parameterized quantum circuit weighting state is reconstructed from the trained parameters and represented as a vector $\bm w=\bm D^{1/2}\bm\alpha$ with normalization before measuring its overlap with $\bm v^\ast$. For \textit{MNIST}, the horizontal axis of Figure~\ref{fig:supp-oos} is truncated for visibility (omitting the largest-budget points), though the full sweeps are used in all computations.

Three observations follow from Figure~\ref{fig:supp-oos}. First, the proposed estimator reaches the accuracy ceiling at small depth ($L=4$ for \textit{Iris}, $L=8$ for \textit{MNIST}), consistent with the under-parameterized regime of the main text. Second, the classical baselines reach the same ceiling as their budgets ($s$ for Nystr\"om, $t$ for Lanczos) grow; on these deliberately small benchmarks the ceiling is high for all methods (test accuracy $\approx 1.00$ on \textit{Iris} and $\approx 0.99$ on \textit{MNIST}, against majority-class baselines of $0.79$ and $0.54$). We therefore read this as parity in clustering quality, not an advantage for the quantum estimator; we make no superiority claim on this perspective.
 
Third, and most importantly for positioning, this perspective is silent about the cost that distinguishes the methods. Both classical baselines can drive eigenvector-approximation quality to the exact solution by spending more of their own resource; what the plot does not show is the kernel-query cost they incur under the quantum fidelity kernel, where - as discussed in Section~\ref{sec:comparison} - each entry must be estimated rather than read at unit cost, whereas our estimator accesses the kernel only through aggregate quadratic forms. In short, our method is preferable not because it recovers a better eigenvector, but because it avoids the per-entry kernel-query cost that this accuracy-only plot leaves out.

\newpage
\subsection{Circuit Depth Comparison}
\begin{figure}[htb!]
    \centering
    \begin{minipage}[t]{0.8\textwidth}
        \includegraphics[width=\textwidth]{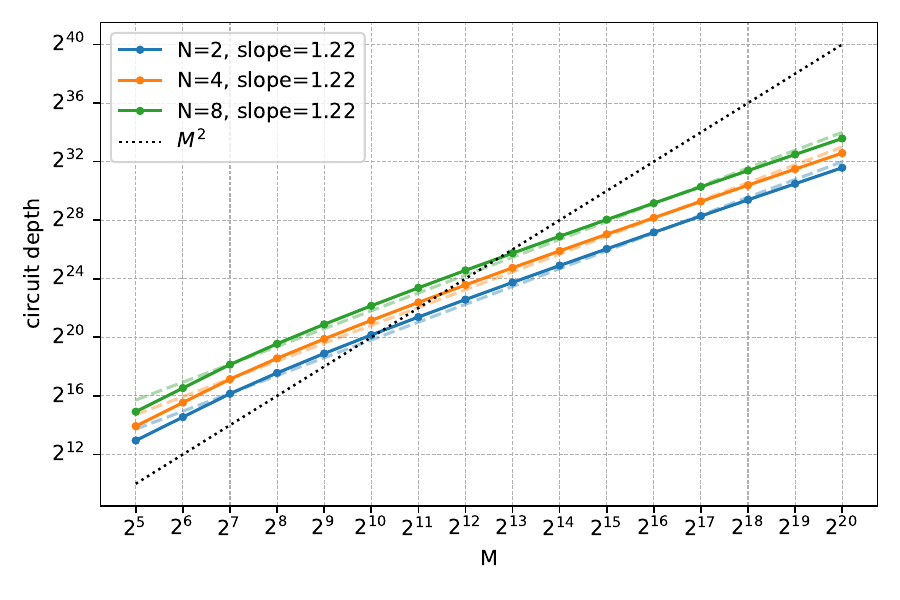}
    \end{minipage}%
    \caption{Depth of the circuit corresponding to $q_a(\alpha)$ under the nearest-neighbor connectivity embedding scheme of Figure~\ref{fig:simulations-embedding (U)} in the main text - the same setup used in the Results section - as a function of exponentially increasing training dataset size, while constraining the gate set to $\{R_x, R_y, R_z, CNOT, CZ\}$. $N=2^n$ denotes the data-register dimension. The colored dashed lines correspond to linear regressions for each trend. The fitted slopes lie below the $M^2$ reference shown as the black dotted line, which corresponds to the naive dense construction of the full kernel matrix.}
    \label{fig:circuit_depth}
\end{figure}

\subsection{Detailed Inference Results of Canonical Datasets}
In general, unsupervised learning models are evaluated using the Adjusted Rand Index (ARI) \cite{RandIndex, ARI}. One key reason for this choice is the ambiguity that arises when interpreting clustering outcomes. However, in binary clustering problems, it is possible to use accuracy as a performance measure—provided that the ARI indicates high reliability, and a majority-vote scheme is applied appropriately. The accuracy results reported in the main text are measured in this manner.
\begin{table}[h!]
    \centering
    \begin{minipage}[t]{0.55\textwidth}
        \centering
        \caption{ARI of \textit{Iris} test dataset}
        \label{tab:ari_iris}
        \begin{tabular}{c c c c c}
            \toprule
            Layers (params.) & Best & Mean & Median & Std \\
            \midrule
            2 (12)  & 1.000 & 0.725 & 0.763 & 0.238 \\
            4 (24)  & 1.000 & 0.944 & 0.948 & 0.047 \\
            6 (36)  & 1.000 & 0.990 & 1.000 & 0.030 \\
            8 (48)  & 1.000 & 1.000 & 1.000 & 0.000 \\
            10 (60) & 1.000 & 1.000 & 1.000 & 0.000 \\
            12 (72) & 1.000 & 0.995 & 1.000 & 0.016 \\
            \bottomrule
        \end{tabular}
    \end{minipage}
    \hfill
    \begin{minipage}[t]{0.55\textwidth}
        \centering
        \caption{Accuracy of \textit{Iris} test dataset}
        \label{tab:acc_iris}
        \begin{tabular}{c c c c c}
            \toprule
            Layers (params.) & Best & Mean & Median & Std \\
            \midrule
            2 (12)  & 1.000 & 0.923 & 0.942 & 0.072 \\
            4 (24)  & 1.000 & 0.987 & 0.988 & 0.011 \\
            6 (36)  & 1.000 & 0.998 & 1.000 & 0.007 \\
            8 (48)  & 1.000 & 1.000 & 1.000 & 0.000 \\
            10 (60) & 1.000 & 1.000 & 1.000 & 0.000 \\
            12 (72) & 1.000 & 0.999 & 1.000 & 0.003 \\
            \bottomrule
        \end{tabular}
    \end{minipage}
    \hfill
    \begin{minipage}[t]{0.55\textwidth}
        \centering
        \caption{ARI of \textit{MNIST} test dataset}
        \label{tab:ari_mnist}
        \begin{tabular}{c c c c c}
            \toprule
            Layers (params.) & Best & Mean & Median & Std \\
            \midrule
            4 (28)   & 0.988 & 0.681 & 0.881 & 0.337 \\
            6 (42)   & 0.981 & 0.885 & 0.961 & 0.160 \\
            8 (56)   & 0.992 & 0.909 & 0.975 & 0.207 \\
            10 (70)  & 0.992 & 0.944 & 0.971 & 0.089 \\
            12 (84)  & 0.988 & 0.969 & 0.971 & 0.014 \\
            14 (98)  & 0.988 & 0.968 & 0.971 & 0.016 \\
            16 (112) & 0.988 & 0.967 & 0.967 & 0.018 \\
            \bottomrule
        \end{tabular}
    \end{minipage}
    \hfill
    \begin{minipage}[t]{0.55\textwidth}
        \centering
        \caption{Accuracy of \textit{MNIST} test dataset}
        \label{tab:acc_mnist}
        \begin{tabular}{c c c c c}
            \toprule
            Layers (params.) & Best & Mean & Median & Std \\
            \midrule
            4 (28)   & 0.997 & 0.891 & 0.969 & 0.132 \\
            6 (42)   & 0.995 & 0.968 & 0.990 & 0.046 \\
            8 (56)   & 0.998 & 0.972 & 0.994 & 0.067 \\
            10 (70)  & 0.998 & 0.985 & 0.993 & 0.025 \\
            12 (84)  & 0.997 & 0.992 & 0.993 & 0.004 \\
            14 (98)  & 0.997 & 0.992 & 0.993 & 0.004 \\
            16 (112) & 0.997 & 0.992 & 0.992 & 0.005 \\
            \bottomrule
        \end{tabular}
    \end{minipage}
\end{table}
\clearpage
\begin{figure}[t!]
    \begin{minipage}{0.5\textwidth}        \includegraphics[width=\textwidth]{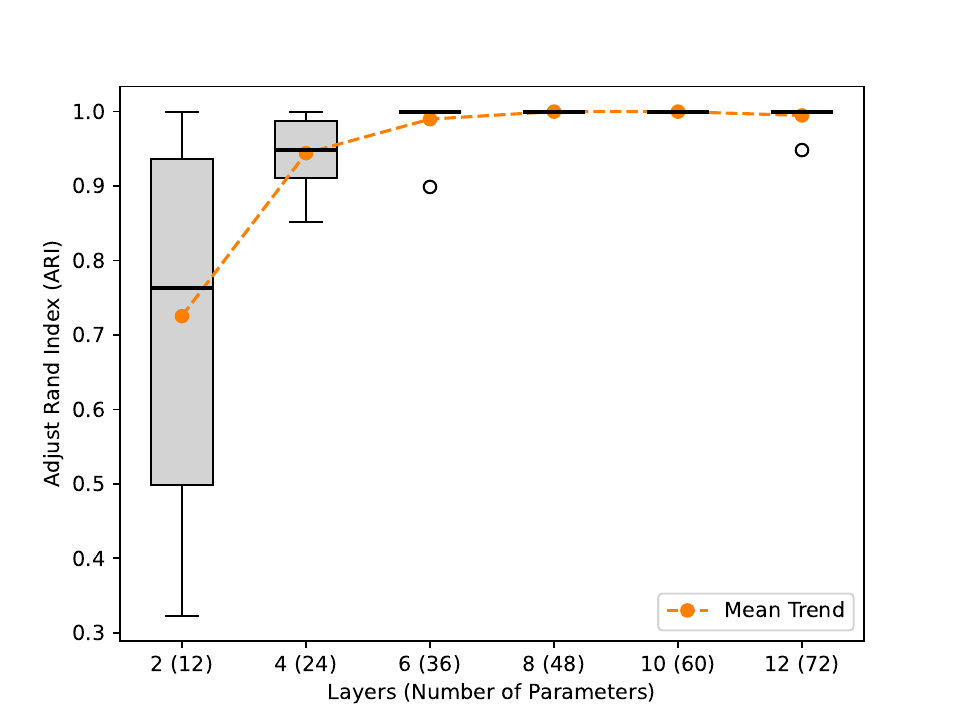}
    \end{minipage}%
    \begin{minipage}{0.5\textwidth}
        \centering
        \includegraphics[width=\textwidth]{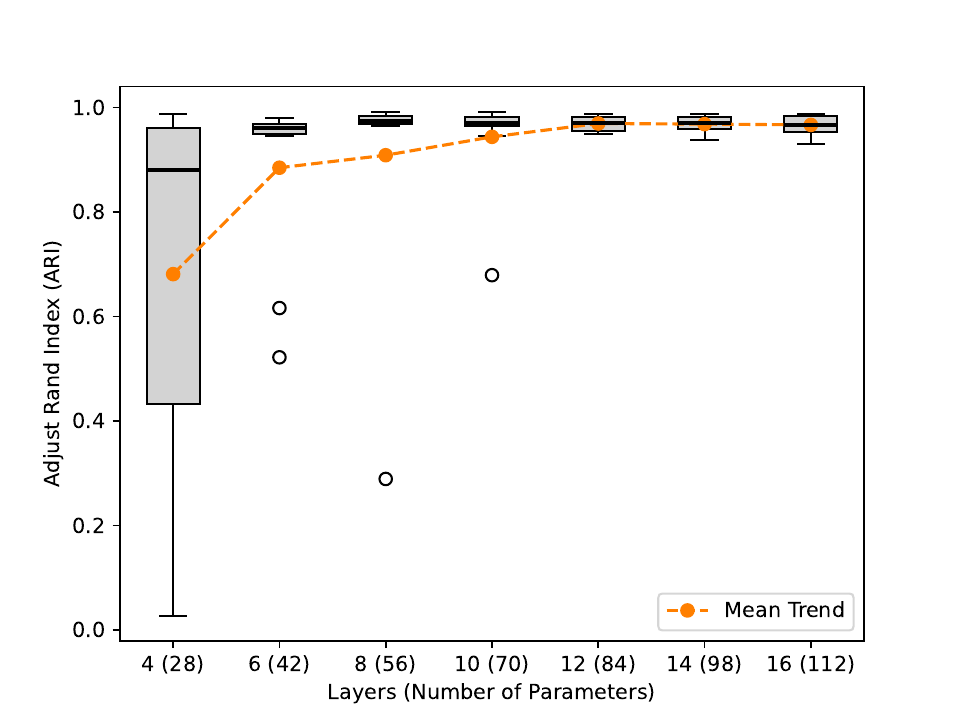}
    \end{minipage}
    \caption{Adjusted Rand Index as a function of the number of layers in the parameterized quantum circuit for the \textit{Iris} dataset (left) and the \textit{MNIST} dataset (right), averaged over 10 different training and test data splits.}
    \label{fig:simulation ari}
\end{figure}

\end{document}